\documentclass[journal]{IEEEtran}

\newcommand{\mytitle}{High-Fidelity State-of-Charge Estimation of Li-Ion Batteries using Machine Learning} 
\newcommand{\myauthor}{Weizhong Wang, Nicholas W. Brady, Chenyao Liao, Youssef A. Fahmy,\\ Ephrem Chemali, Alan C. West, and Matthias Preindl} 

\usepackage{ifthen}
\usepackage[a-1b]{pdfx}

\usepackage{lmodern}

\usepackage{amsmath} 
\usepackage{amsthm}

\usepackage{microtype} 
\usepackage[hyperref]{xcolor}

\usepackage{graphicx}
\usepackage{epstopdf}
\usepackage{pdfpages}
\usepackage{booktabs} 
\usepackage[inline,shortlabels]{enumitem}

\usepackage{url}
\usepackage{hyperxmp}
\usepackage{hyperref}
\usepackage{bookmark}

\hypersetup{
	unicode,
	bookmarksopen,
	bookmarksopenlevel=0,
    colorlinks, 
	linkcolor={black}, 
    anchorcolor={black}, 
	citecolor={black}, 
	filecolor={black}, 
	menucolor={black}, 
	runcolor={black}, 
	urlcolor={black}, 
	baseurl = {http://www.matthiaspreindl.com},
	pdftrapped = {False}
}

\usepackage{polyglossia}
\setdefaultlanguage{english}
\addto\captionsenglish{} 
\usepackage{csquotes}
\usepackage[
    backend=biber, texencoding=utf8, 
    style=ieee,
	minnames=1, maxnames=2,
    hyperref=true,
	doi=false,
	url=false, 
	isbn=false
]{biblatex}
\usepackage{subfigure}
\usepackage{soul} 
\usepackage{csquotes}
\usepackage[
	backend=biber, texencoding=utf8, 
	style=ieee,
	hyperref=true,
	doi=false,
	url=false, 
	isbn=false
]{biblatex}

\usepackage{chemformula} 

\graphicspath{{./graphic/}}
\DeclareGraphicsExtensions{.png,.pdf}

\begin{document}

\title{\mytitle} 

\author{
	\myauthor 

	\thanks{
		%
		Weizhong Wang, Chenyao Liao, Youssef A. Fahmy, and Matthias Preindl are with the Department of Electrical Engineering, Columbia University in the City of New York, New York, NY 10027, USA (e-mail: ww2427@columbia.edu; cl3757@columbia.edu; yaf2105@columbia.edu; matthias.preindl@gmail.com). 
		
		Nicholas W. Brady and Alan C. West are with the Department of Chemical Engineering, Columbia University in the City of New York, New York, NY 10027, USA (e-mail: nwb2112@columbia.edu; acw7@columbia.edu). 
		
		Ephrem Chemali is with the Department of Electrical and Computer Engineering, McMaster University, ON L8S 4K1, Canada (e-mail: ephrem.chemali@gmail.com)
	}
}

\maketitle

\begin{abstract}
This paper proposes a way to augment the existing machine learning algorithm applied to state-of-charge estimation by introducing a form of pulse injection to the running battery cells. 
It is believed that the information contained in the pulse responses can be interpreted by a machine learning algorithm whereas other techniques are difficult to decode due to the nonlinearity. 
The sensitivity analysis of the amplitude of the current pulse is given through simulation, allowing the researchers to select the appropriate current level with respect to the desired accuracy improvement. 
A multi-layer feedforward neural networks is trained to acquire the nonlinear relationship between the pulse train and the ground-truth SoC.
The experimental data is trained and the results are shown to be promising with less than 2\% SoC estimation error using layer sizes in the range of 10 - 10,000 trained in 0 - 1 million epochs. 
The testing procedure specifically designed for the proposed technique is explained and provided.
The implementation of the proposed strategy is also discussed. The detailed system layout to perform the augmented SoC estimation integrated in the existing active balancing hardware has also been given.
\end{abstract}



\definecolor{limegreen}{rgb}{0.2, 0.8, 0.2}
\definecolor{forestgreen}{rgb}{0.13, 0.55, 0.13}
\definecolor{greenhtml}{rgb}{0.0, 0.5, 0.0}

\section{Introduction}



\IEEEPARstart{A}{ccurate} state-of-charge (SoC) estimation is necessary for optimal battery management and safe and reliable utilization of battery powered devices, such as electric vehicles (EVs) and grid level energy storage. For lithium-ion batteries, in particular, SoC estimation is difficult because the relationship between the SoC and the open-circuit voltage (OCV) is non-linear, as can be seen in Fig. \ref{fig:OCP_SoC}. In certain ranges of the SoC in Fig. \ref{fig:OCP_SoC}, the voltage is completely flat with respect to the SoC due to phase changes occurring within the system; this makes it challenging to estimate the SoC from voltage measurements. A variety of methods have been proposed to estimate the SoC in lithium-ion batteries.

Unlike the fuel level in traditional combustion engine vehicles, the SoC cannot be directly measured in EV applications. However, the SoC is internally linked with direct measurement (voltage, current, temperature and capacity) and can be extracted by using battery intrinsic relations and/or control theory. 

\subsubsection{Open-Circuit Voltage (OCV) mapping}
The techniques of estimating the SoC have been extensively investigated. The most straightforward method is to map the OCV to the SoC, as a one-to-one translation can be found between SoC and OCV under certain conditions. Given a specific OCV, the corresponding SoC can be accurately interpreted if the measured condition matches the one where the OCV-SoC map is acquired. In other words, the OCV-SoC map varies with the testing conditions, such as temperature and aging status, which introduces a significant amount of variability and can bias the SoC estimation \cite{Chemali2016,Waag2014,Baronti2011}. Even the direction of current flow (charging/discharging) will affect the OCV-SoC map significantly according to \cite{Roscher2011}. In addition, complete electrochemical equilibrium cannot be achieved within a short time frame \cite{Waag2013c}. Therefore, while the battery is under load, it is unfeasible to perform real-time updates of the SoC based on OCV measurements. For these reasons, OCV-based SoC estimation is commonly used as a complementary or corrective method running in the background \cite{Xiong2017}. \looseness = -1

\subsubsection{Coulomb-Counting}
Coulomb-Counting identifies a SoC estimation technique that integrates the battery current, i.e.\ counts the Coulombs. Hence, it can identify an SoC difference but requires knowledge of an initial SoC value, which can be obtained with an OCV-SoC map in a well known condition. Coulomb counting (or Ah counting) integrates the current passed in/out of the battery with respect to time and converts it to the SoC using the following expression:

\begin{equation}
    \mathrm{SoC} = \mathrm{SoC_0} + \eta \int \frac{i}{C} dt
    \label{eq:soc}
\end{equation}
where $\mathrm{SoC_0}$ is the initial state of the SoC; $C$ is the present capacity of the cell. The charging/discharging efficiency is denoted as $\eta$. The current that charges/discharges the battery is $i$. However, the accuracy of SoC estimation would be compromised if low-res current sensors are used or the capacity is not updated as the battery ages \cite{Ng2009,Baronti2011}. Especially in situations where the SoC cannot be regularly corrected by OCV-based methods, the predicted SoC significantly drifts away from the true value and misleads other functions in BMS. As a result, coulomb counting is commonly used in the laboratory environment where the aforementioned uncertainties can be reasonably controlled. 

\subsubsection{Model-based Observer}
To reduce the uncertainties of the open-loop SoC estimation methods mentioned previously, techniques with feedback mechanisms to correct for possible bias and real-world compromises (such as sensor resolution) have been extensively investigated. Modern nonlinear state estimators and observers are commonly adopted. Particularly, Kalman-Filter (KF) based technologies \cite{Plett2004c,Wang2016c,Dai2012,Sepasi2014,Xiong2017,Plett2005}, recursive least square methods (RLS) \cite{He2012,Xia2018}, and slide-mode observers \cite{Ning2016,Belhani2013,Liu2016} have been heavily researched as they provide reasonable estimation accuracy and relatively robust performance. 

However, constructing such an observer requires precise system modeling \cite{He2012} for the specific type of battery in the system and repetitive hand tuning to select a well-behaved covariance matrix. As the battery ages, the derived battery model using a `fresher' cell's data is biased and may even be invalid. The capacity decreases while impedance increases for aged cells, which can result in an offset/error from the true SoC and even divergence of the observer. In addition, the initial states of the observer that are fed from external sources significantly affect the performance of the estimator, in terms of convergence and accuracy. 

\subsubsection{Data-driven methods}
With advancements in computation and an abundance of real world data, machine learning or specifically neural network-based methods are providing researchers with the ability to achieve significant advancements in many fields \cite{Kri2012,ciresan2012,Hinton2012,Ma2015,Wang2016}. SoC estimation applying neural network-based methods has also drawn attention \cite{Chemali2017,Chemali2018,Du2014,Charkhgard2010}. 

Compared to a 2\% average SoC error achieved by model-based observers \cite{Plett2004c,Wang2016c,Dai2012,Sepasi2014,Xiong2017,Plett2005}, 4\% RMS error on terminal voltage is achieved with 2-layer neural network and 30 neurons in the hidden layer \cite{Charkhgard2010}. However, should further error reduction be desired, neural networks need the help of external filtering/observer (like Kalman-filter in \cite{Du2014}). \citeauthor{Chemali2018} directly mapped the measurements of the cell (instantaneous and average terminal voltage, temperature, and average current) to the SOC estimation and is able to achieve a mean absolute error below 1\% \cite{Chemali2018}. This paper also showed that the number of layers and neurons had a minimal effect on the SOC estimation accuracy – a 2-layer network with 2 neurons per layer seems to be a good compromise between computational time and estimation accuracy. Nevertheless, the worst-case error through the entire test was as high as 7\%.

In cases where estimation performance is limited by an OCV-SoC plateau, seen in Fig. \ref{fig:OCP_SoC}, or when complete equilibrium of the battery is unfeasible, additional information can be gathered by injecting an augmented current profile into a battery under load. 
This paper hypothesizes that passing current pulses through a battery and measuring the voltage response to these pulses can be used to retrieve information about a lithium-ion battery's SoC. Because these measured electrochemical responses do not have an obvious relationship with the SoC, a neural-network can be used to learn the relationship and reconstruct the information. 

This paper is organized as follows. In Section \ref{sec:simulation}, a previously constructed electrochemical model is used to prove the concept and provide insights of correlation between pulse amplitude and accuracy improvement. A neural network constructed using TensorFlow was tested in Section \ref{sec:ml} whether OCV measurements and current pulse information could effectively reconstruct the SoC information of the system. The testing procedure and experimental results using NMC cells captured by real-time battery testing system are shown in Section~\ref{sec:data}. The hardware and software are being developed for practical implementation of the pulse derived SoC estimation in Section \ref{sec:pack}. The results of the paper are summarized and proposed future work is discussed in Section \ref{sec:Conclusion}.




\section{Hypothesis and Proof-of-Concept Simulation}
\label{sec:simulation}
%

\begin{figure}[t!]
	\centering
	\includegraphics[width=0.45\textwidth]{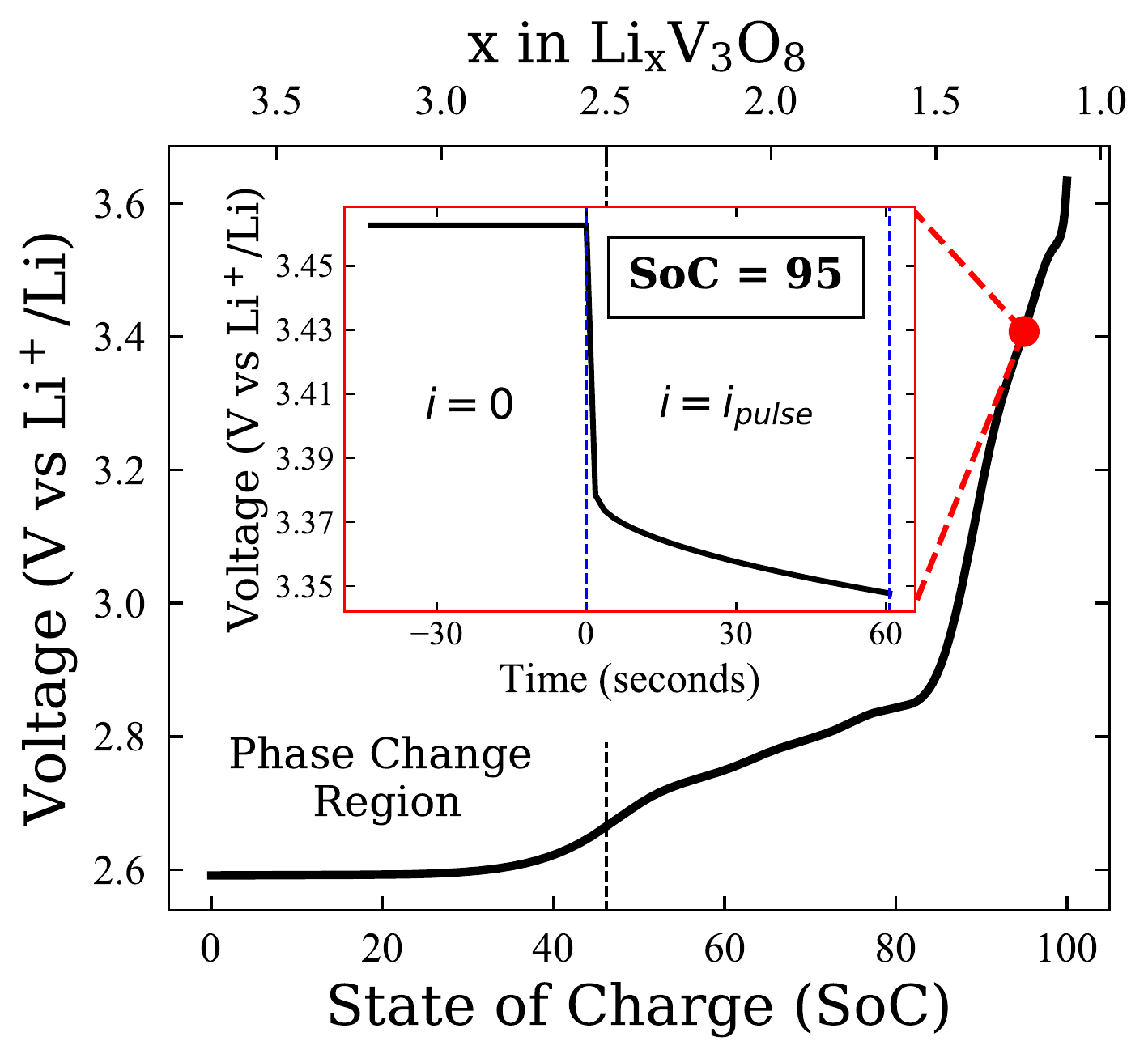}
	\caption{
		The simulated open-circuit voltage (OCV) of the lithium trivanadate (\ch{Li_{x}V3O8}) electrode vs lithium metal as a function of state of charge (SoC). The inset is a representation of how the pulse is implemented for the system; this pulse was taken at SoC $= 95$.
	}
	\label{fig:OCP_SoC}
\end{figure}

The numerical details for the implementation of the \ch{Li_{x}V3O8} electrochemical model are found in refs. \cite{Brady2016,Brady2018}. For the current pulses done in this paper, the cathode thickness was assumed to be 500 $\mathrm{\mu m}$, the porosity was assumed to be 0.45, with the volume fraction of active material (\ch{Li_{x}V3O8}) being 0.48, and the volume fraction of conductive material being 0.07, and the crystal size of the active material was assumed to be 120 nm in the [001] direction. 

\begin{figure}[h!]
	\centering
	\includegraphics[width=0.46\textwidth]{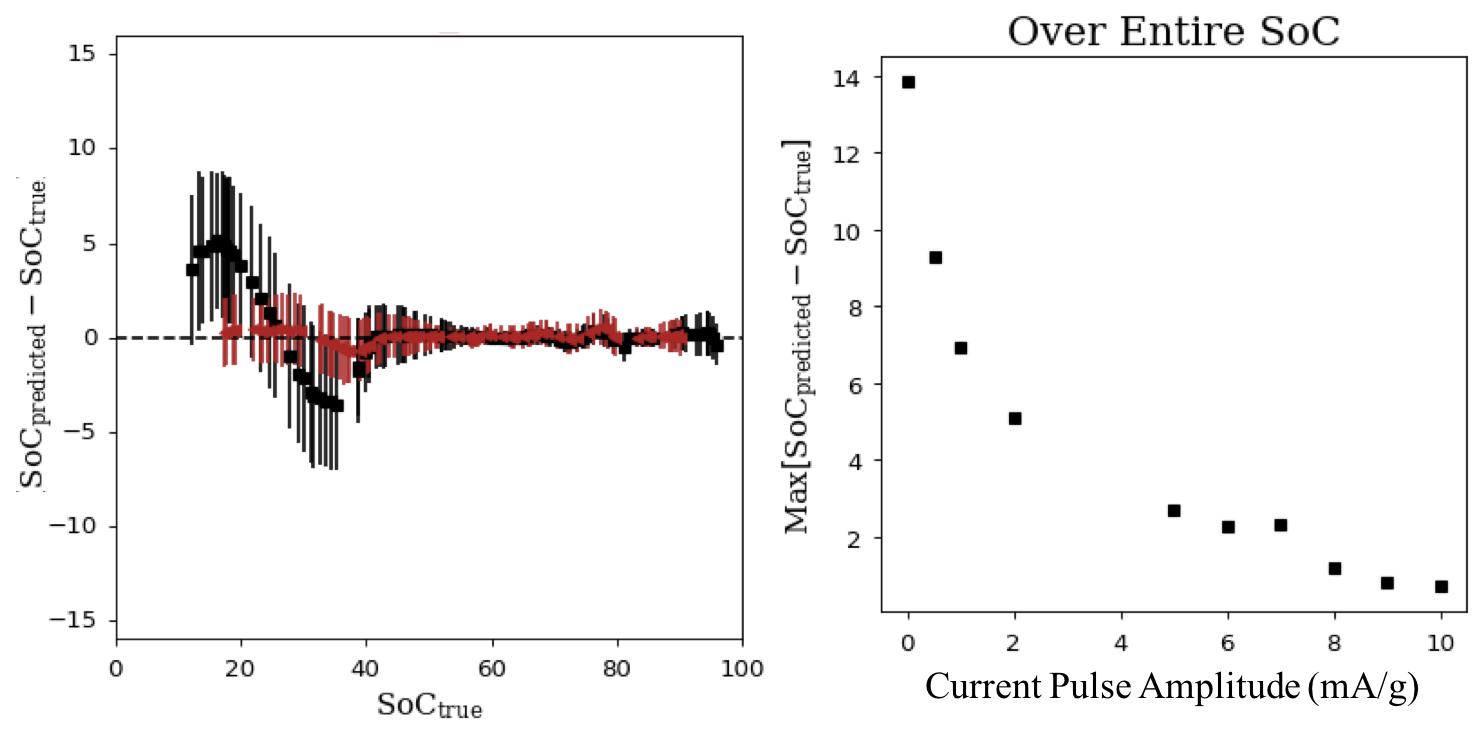}
	\caption{
	    (Left) Deviation (and uncertainty) of the predicted SoC from their true value as a function of the true SoC. Estimates were produced from the neural network using only the open-circuit voltage data (black) and using a current pulse of $10\times$ (brown). (Right) Maximum error of the SoC estimation produced from varying current pulses; current pulse of 0 is the estimation using OCV data and every mA/g is C/20.
	}
	\label{fig:OCP_Curr_10_with_Summary}
\end{figure}

Fig. \ref{fig:OCP_SoC} shows how the OCV of \ch{Li_{x}V3O8} varies as a function of SoC. It is observed that the relationship between the SoC and the OCV is non-linear and in particular, when this material goes through a phase-change (approximately $\mathrm{SoC} \leq 40$), the OCV is constant, i.e. $ \mathrm{OCV} \neq f(\mathrm{SoC})$.
Because of the non-linearity and because of the OCV-SoC plateau, it is difficult to estimate the SoC from OCV measurements alone. Fig. \ref{fig:OCP_Curr_10_with_Summary} (left - black data) shows how the estimates produced from the OCV (black) deviate from the true SoC. The OCV derived estimates are precise and accurate in the range $60 \le \mathrm{SoC} \le 100$, but are imprecise and inaccurate in the range $\mathrm{SoC} < 60$. 

To gain more information about the battery and thereby obtain better estimates of the SoC, pulses were constructed by lithiating the \ch{Li_{x}V3O8} cathode at a current rate of C/18, allowing the system to rest for 2 hours, then passing a pulse current at various amplitudes and measuring the potential for 60 seconds at a sampling rate of 1 Hz. An example of the voltage measurements derived from one of these current pulses can be seen in the inset of Fig. \ref{fig:OCP_SoC}. Fig. \ref{fig:OCP_Curr_10_with_Summary} (left - brown data) shows that the estimates derived from voltage measurements during a current pulse are more accurate and precise than the estimates from the OCV measurements, especially in the range $\mathrm{SoC} < 60$.

Fig. 2 (right) shows the relationship between accuracy of the SoC estimates and the amplitude of the pulse current. A pulse current of 0 corresponds to the estimate obtained using the OCV data. It is observed that as the amplitude of the injected current increases, the accuracy of the estimation improves and this improvement is most pronounced at low amplitudes. Additionally, the improvement appears to plateau at $8\times$. If an objective function seeks to minimize estimation error, while also minimizing the cost incurred by pulsing the system, the relationship observed in Fig. \ref{fig:OCP_Curr_10_with_Summary} (right) implies that there is an optimal current to apply for this particular system. It should be noted that there is a preference toward low amplitude pulses especially in EV applications because low amplitude pulses are easier to implement through the balancing hardware.

\section{Test Cases Design}
\label{sec:data}
  Injecting pulses can clearly improve the robustness of SoC estimation as shown in the previous section. To build up the initial machine learning model and implement the concept in real battery systems, the pulse data sets that are used to train the neural network are first obtained in the laboratory environment. Therefore, a systematic testing procedure is proposed to ease the procedures that future researchers need to go through. The test cases basically follow the standard hybrid power pulse characterization (HPPC) test regulated by DOE \cite{Idaho2010} but with distinguished modifications to adapt the optimized pulse amplitude as well as the facilitated aging tests, as discussed in Section \ref{sec:simulation}.   

\subsection{Testbench} 
The lithium nickel manganese cobalt oxide (NMC) cells are selected initially to explore the possibility of implementing machine learning. Compared with other chemistries, such as lithium cobalt oxide (LCO), NMC provides higher boost current and longer life-span and therefore is commonly used in automotive and energy storage systems \cite{BatteryU}.

\begin{figure}[]
    \centering
    \includegraphics[width=0.35\textwidth]{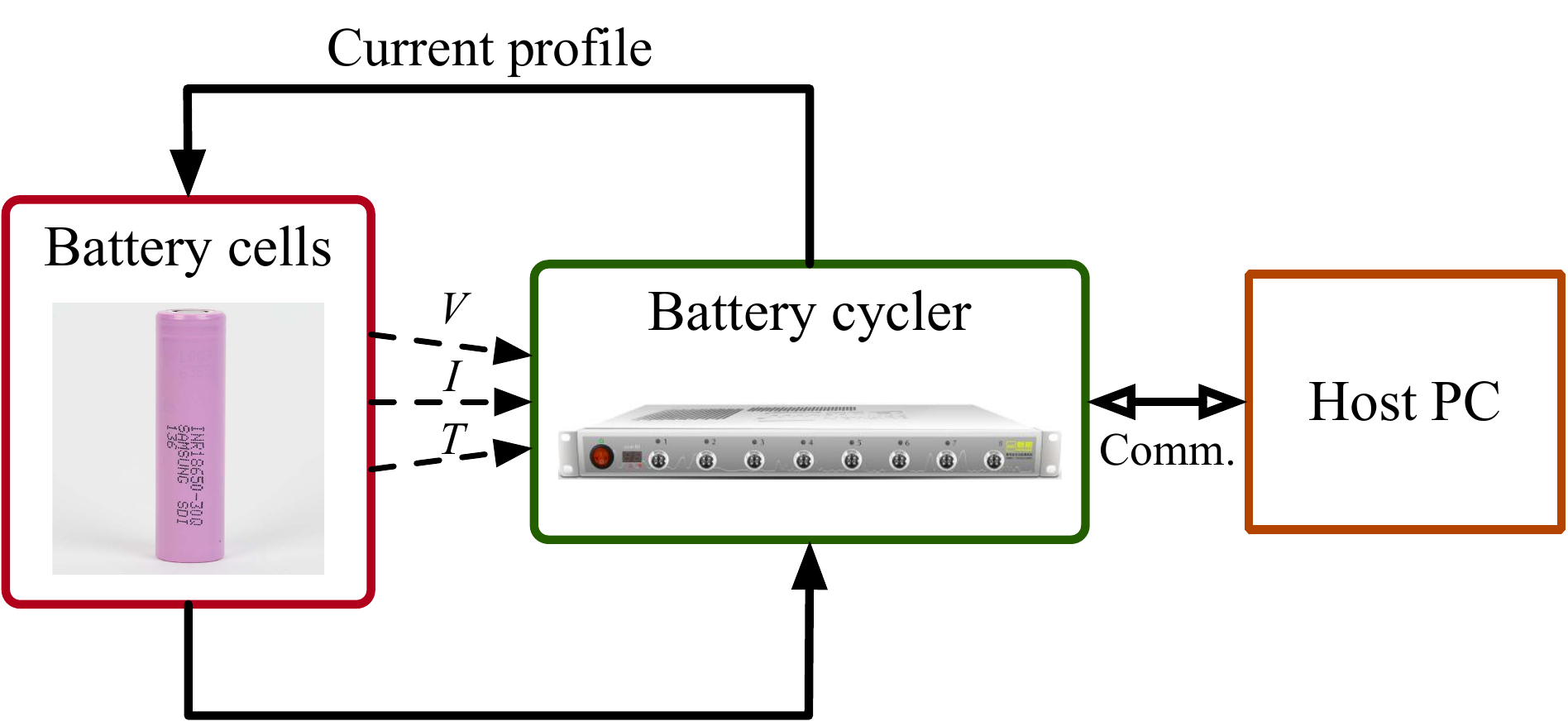}
    \caption{The overview of the testing and data acquisition system}
    \label{fig:testsetup}
\end{figure}

The testbench consists of (i) real-time battery cycler with thermal couples, Neware BTS4000 series; (ii) host PC recording and uploading data to data base; (iii) NMC high-energy cells (parameters shown in Table \ref{tab:cell_para}). The complete system is shown in Fig. \ref{fig:testsetup}. \looseness = -1

\begin{table}[b!]
    \centering
    \caption{Cell parameters}
    \begin{tabular}{lr}
        \hline
        Cell chemistry   & NMC \\
        Nominal capacity & 3000 mAh\\
        Cut-off voltage/current & 2.5 V/150 mA\\
        Maximum voltage & 4.2 V\\
        Maximum charging/discharging current & 4/15 A\\
        \hline
    \end{tabular}
    \label{tab:cell_para}
\end{table}

The real-time battery cycler is capable of testing the cells using constant current (CC), constant voltage (CV), CCCV, and dynamic current profiles (driving cycles) with a sampling resolution of 0.1s. The hi-res measurements, including voltage, current and temperature, are uploaded to the database through the communication line with the host PC for later data process.

\subsection{Capacity Check}
Before discussing the pulse train structure, the capacity should be checked regularly in order to interpret the SoC correctly. One of the most obvious consequences that can be observed when the battery cells are aged is the capacity fade. In addition to the uncertainties of the estimator, if the capacity is also outdated, it further degrades the accuracy of SoC estimation as the SoC and present capacity are interacting with each other according to the SoC definition in Eq. (\ref{eq:soc}). Note that in this paper the sign convention for the current is positive for charging and negative for discharging.

\begin{figure}
    \centering
    \subfigure[]{
        \includegraphics[height=0.18\textwidth]{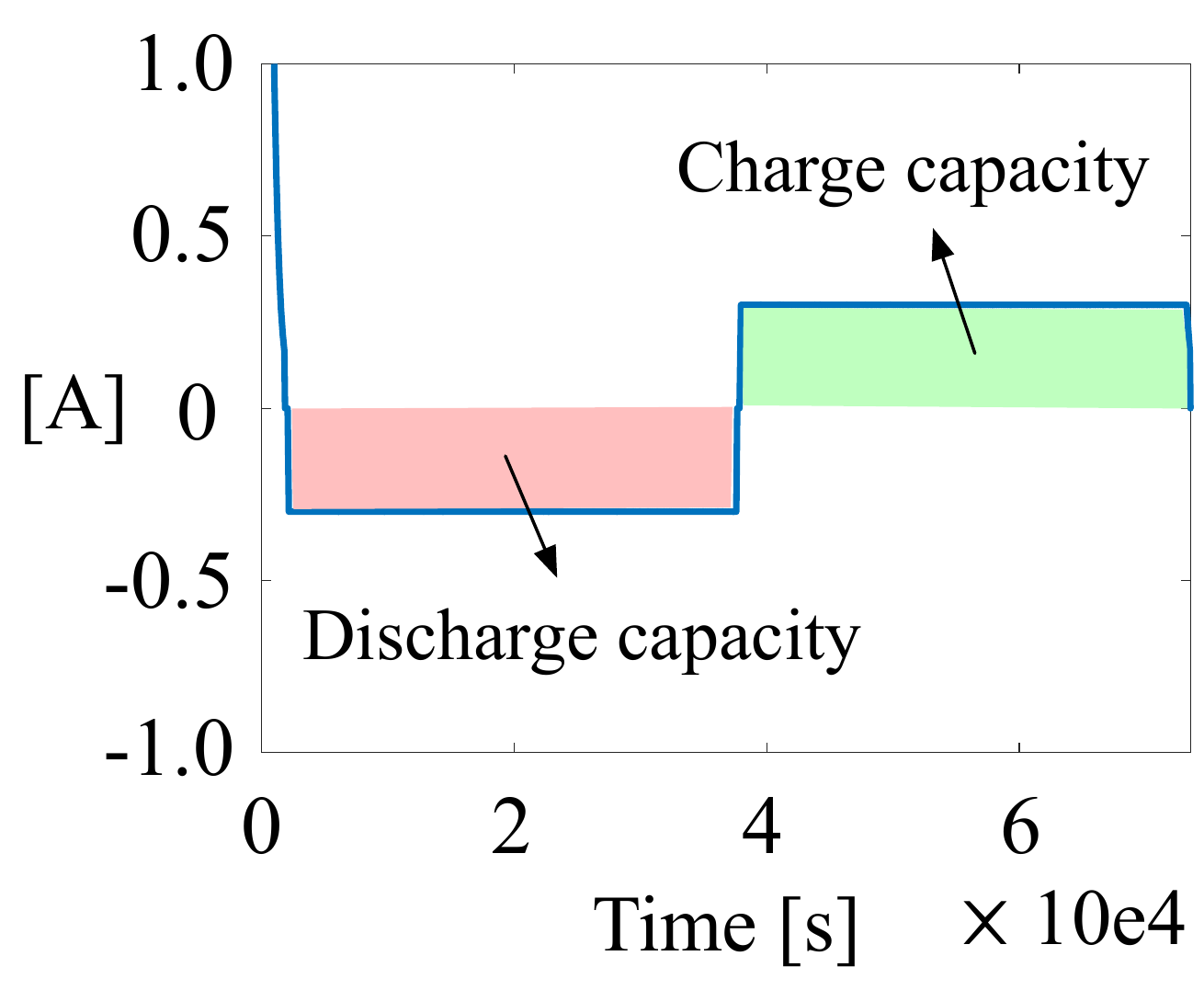}
        \label{fig:I_capcheck}
    } 
    \subfigure[]{
        \includegraphics[height=0.17\textwidth]{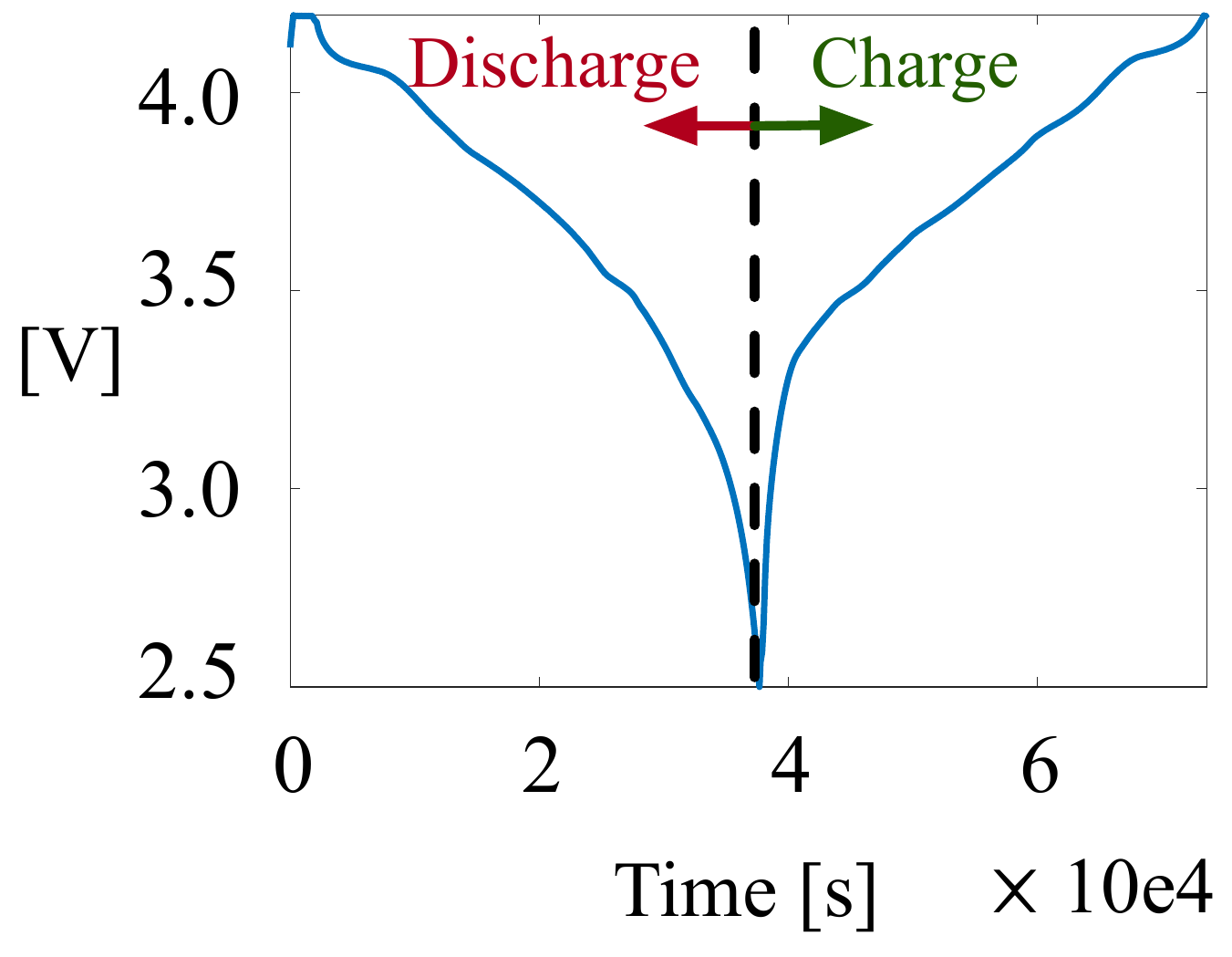}
        \label{fig:V_capcheck}
    }
    \caption{Capacity check: (a) Current, (b) Voltage}
    \label{fig:cap_check}
\end{figure}

Should the accurate SoC be desired, knowing the capacity in advance is essential to train the machine learning algorithm. Capacity is often defined with the current level. Higher current (either charging or discharging) will result in lower capacity due to the internal resistance of the battery \cite{Wang2016b}. However, in order to obtain the approximately true capacity that is available in the cell, a relatively small current rate (0.1 C-rate) is trickled in to fully charge and discharge the cell. The smaller the current rate is, the better the internal resistance can be ignored. Both charge and discharge capacity can be obtained by integrating the corresponding parts, as shown in Fig. \ref{fig:cap_check}. Note that, this capacity should be used for back-calculating the precise SoC breakpoints in the following section. 

\subsection{Pulse Train}
The pulse train needs to be carefully designed in order to maximize the information that can be extracted from limited data points. As discussed previously, the properties of the pulse determine the accuracy of the SoC estimation. It can be seen that higher current level contributes to lower estimation error. However, the feasibility of the current amplitude in the real battery system needs to be investigated. So a trade-off should be made between accuracy and feasibility. Especially in real-life EV systems, the pulses should not interrupt how the drivers drive or leave any obvious sign that the BMS is trying to reconstruct the SoC. The violation, for example but not limited to, can be unexpected acceleration. But it can be as 'stealthy' as a current sharing between cells when the battery needs balancing, which can be achieved by active balancing topology describe in the next section. In this paper, the current amplitude is chosen to be 1 C-rate since it potentially will decrease the error more while keeping the cells away from maximum allowed current.

\begin{figure}[h!]
    \centering
    \includegraphics[height=0.3\textwidth]{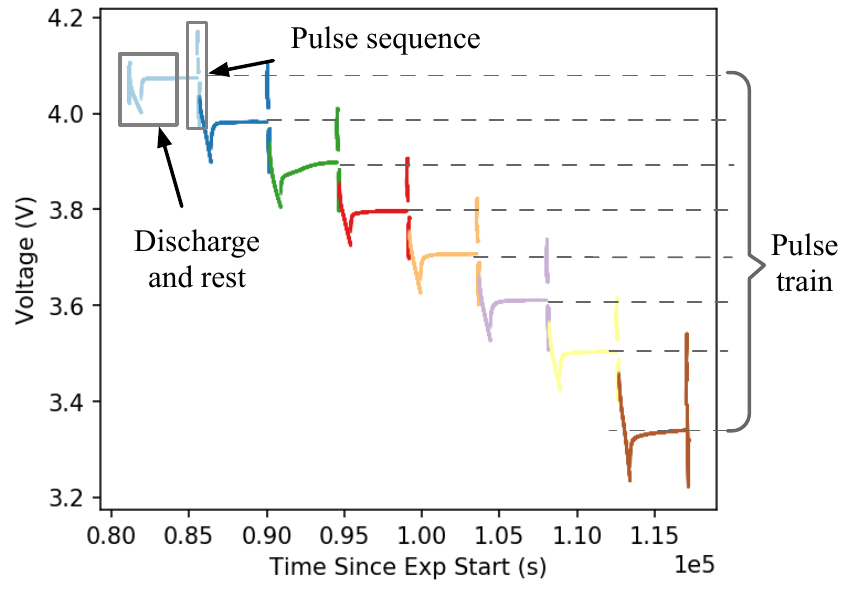}
    \caption{Pulse train with intermediate discharging and rest}
    \label{fig:HPPC_exp}
\end{figure}

The pulses are injected at every 10\% SoC. Finer resolution can also be achieved but with the compromise of testing time. Firstly, the battery cells are fully charged by CCCV, followed by 1 - 2-hour rest to allow complete equilibrium inside of the battery. The battery is then discharged at 1 C-rate to 90\% SoC. By allowing 1-hour relaxation before injecting pulses, the voltage response isolates the charge-transfer and/or charge diffusion effects that are induced by previous current excitation. The subsequent voltage response will be purely excited by the current pulses. If the cell is not well rested, the charge history will be coupled into the pulse response, which makes the results less accurate. A 1-min long charge pulse and discharge pulse with 1-min rest between them are injected at 90\% SoC. Then the cell is discharged to 80\% SoC and repeat the same sequence as it is for 90\% SoC. This procedure keeps repeating until cut-off voltage is reached at any time. The sample results for testing sequence is shown in Fig. \ref{fig:HPPC_exp}.

\begin{figure}[h!]
    \centering
    \includegraphics[height=0.3\textwidth]{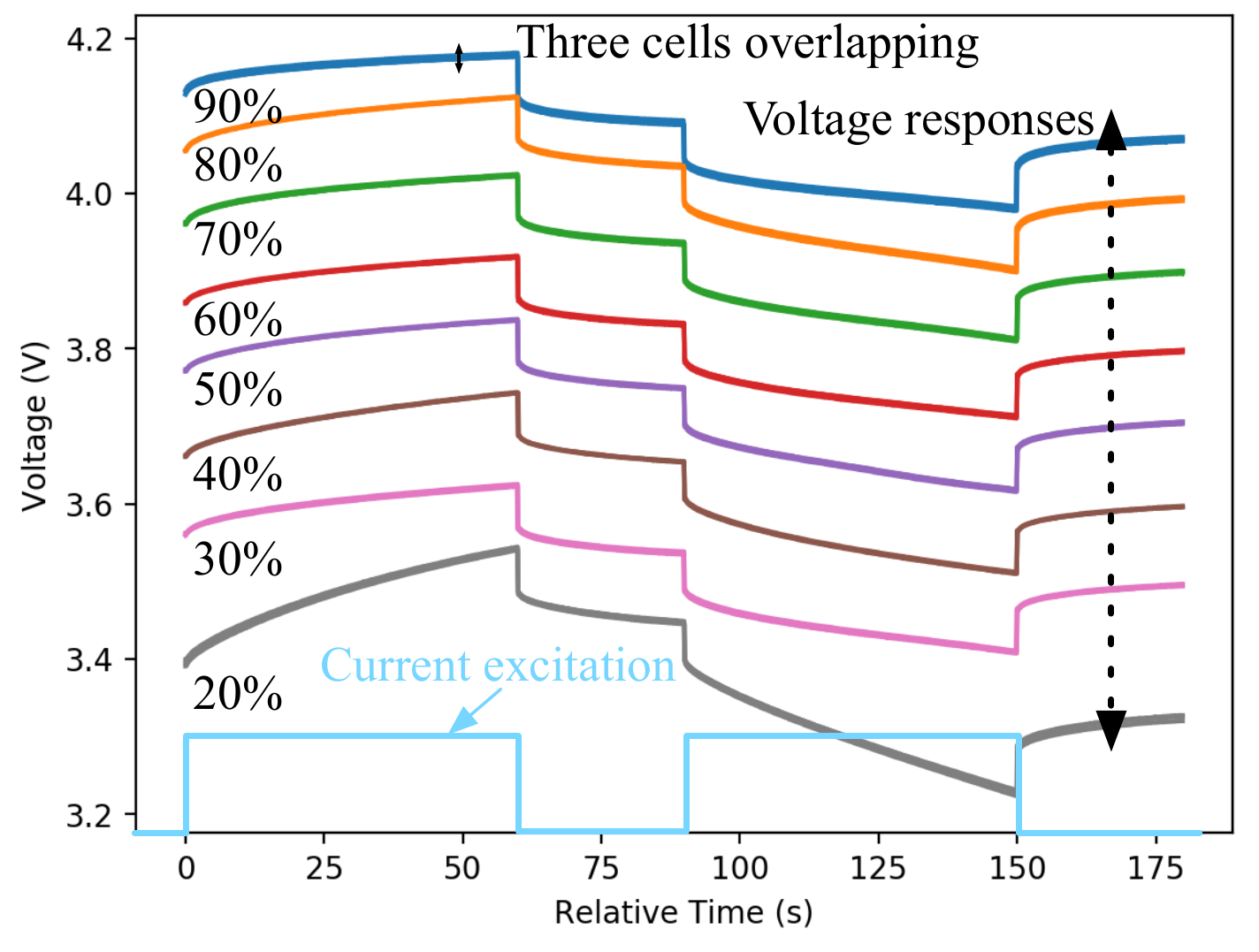}
    \caption{Supperimposed cell voltage responses between 20\% and 90\% SoC from three cells}
    \label{fig:Fresh_90_20_pulse}
\end{figure}

Cells occasionally show strong individuality in terms of aging trends and responses to current due to the variations in manufacturing processes. Three cells are tested under same conditions as a batch to minimize the individuality by comparing and averaging the resulting data. Fig. \ref{fig:Fresh_90_20_pulse} illustrates the voltage responses excited by the current pulses from 90\% to 20\% SoC levels. The major difference among them is the voltage level where they operate gradually decreases as the cells deplete more. There are also more subtle differences which hardly can be differentiated by bare eyes but can be captured by the machine learning algorithm, for example higher voltage drops when current just applies to the cells as cells discharge. At each SoC breakpoint, the results from three cells are superimposed on each other. It shows high consistency across the entire SoC test points. 

\subsection{Overall Test Procedure}
As the battery ages, the performance of the model-based SoC estimators significantly degrade as they highly rely on an accurate model, especially on capacity as expressed in Eq. (\ref{eq:soc}). Normally, a joint estimator or a separate slow-react estimator needs to be added for capacity estimation \cite{Wang2015,Zou}, which inevitably increases the complexity of the BMS. 

Aging a battery is a time-consuming task. To accelerate the aging process, a pre-defined aging procedure is proposed here. The cells under test fully discharge with a CC at 1 C-rate, and followed by fully charge with a CC at 1 C-rate to maximum voltage and CV until current drops below 150 mA. The aging test will be terminated when the capacity reaches 80\% of its original one, which is normally called end-of-life (EOL) for EV application.

\begin{figure}[]
    \centering
    \subfigure[]{\includegraphics[width=0.48\textwidth]{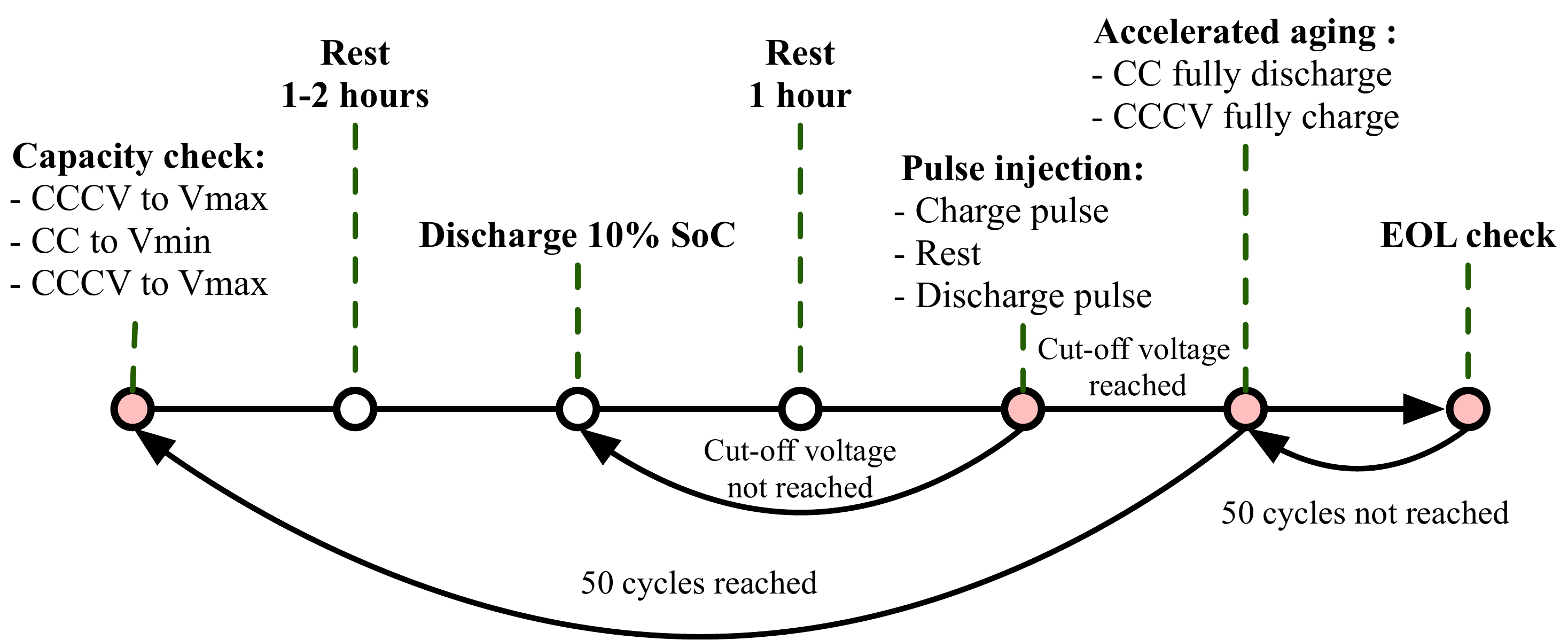}}
    \label{fig:test_flowchart}
    
    \subfigure[]{\includegraphics[height=0.35\textwidth]{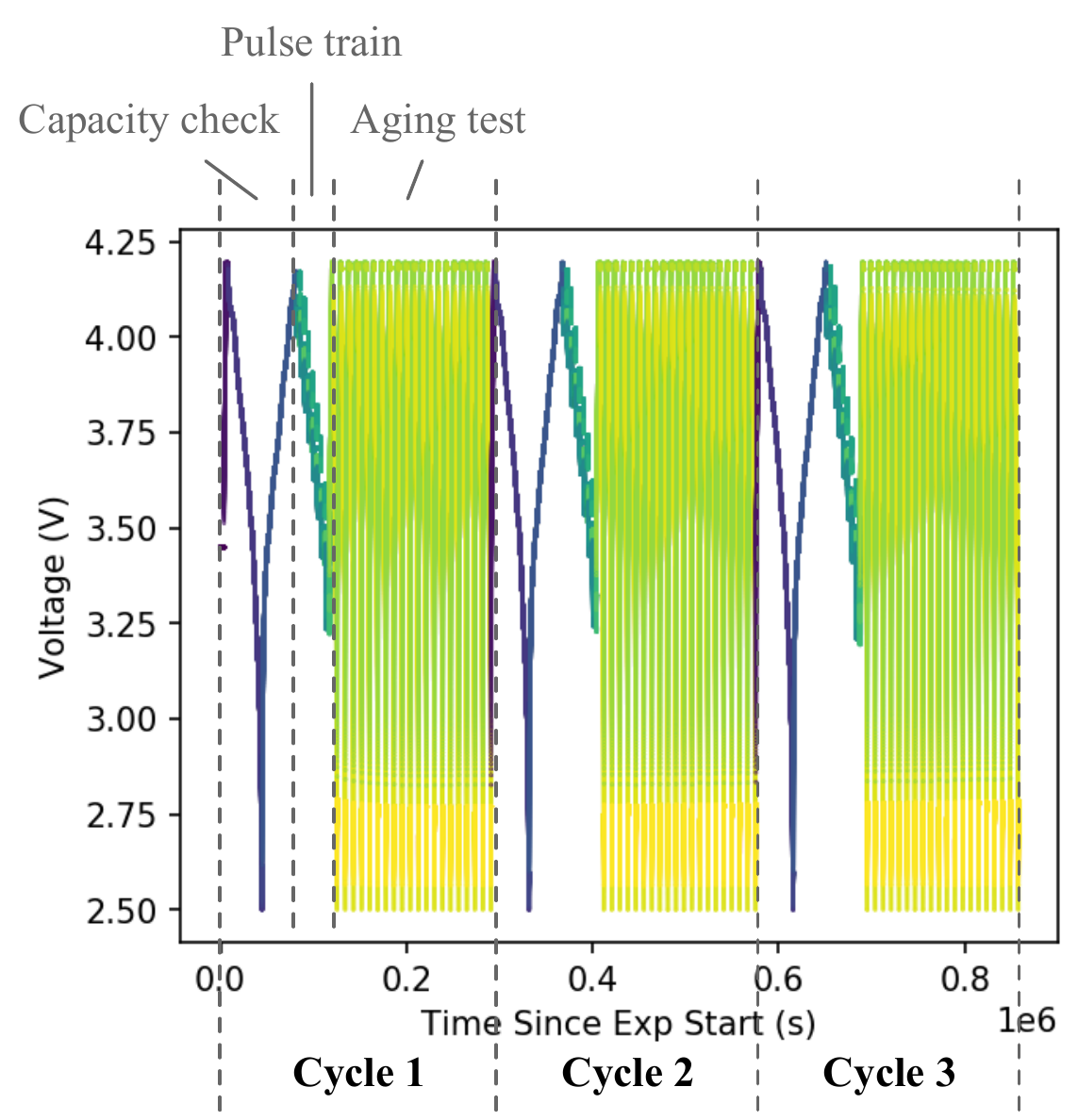}}
    \label{fig:Full_test_exp}
    
    \caption{The test procedure: (a) The flowchart, (b)Complete test sequence captured by the battery cycler}
    \label{fig:Test_procedure}
\end{figure}

Combining the capacity check, pulse train and accelerated aging test completes the testing procedure design. The entire test procedure is summarized in Fig. \ref{fig:Test_procedure}.


\section{Machine Learning Finding the Correlation}
\label{sec:ml}

\begin{figure}[h!]
    \centering
    \includegraphics[width=0.45\textwidth]{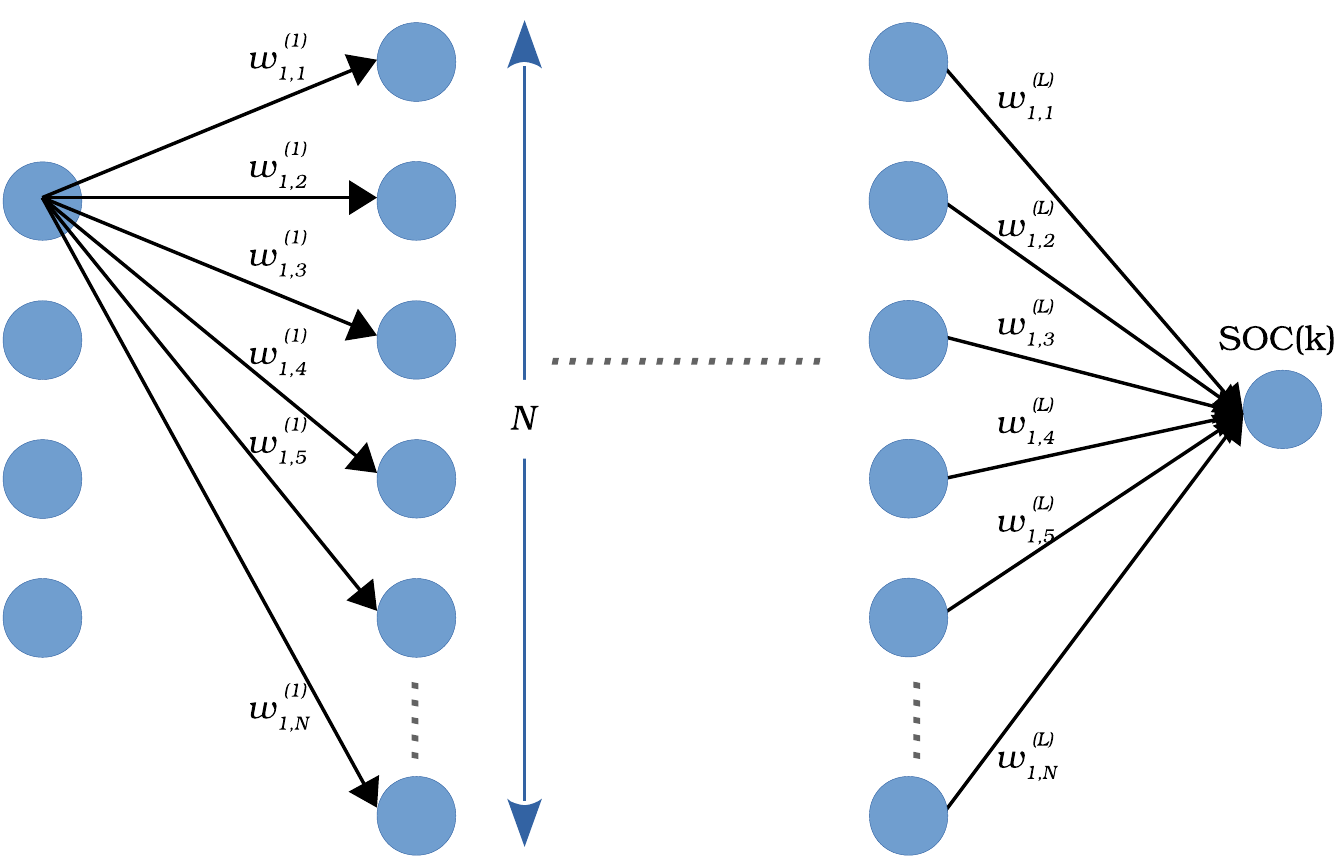}
    \caption{Architecture of Feedforward Neural Network (FNN).  The input data is the vector representing the recorded pulse-train and the output of the FNN is the estimated SoC for pulse-train $p$.}
    \label{fig:FNN}
\end{figure}

\sloppy Feedforward neural networks (FNN), shown in Figure \ref{fig:FNN} to have 2-layer and multi-layer architectures, can in principle, model most non-linear systems by mapping inputs to a desired output.  

In this paper, the pulse-train generated from previous tests is fed as an input to an FNN and an estimated SoC is provided as an output of the network.  The network is trained by computing the difference between this estimated value and the ground-truth or ideal SoC values.  Therefore, a typical input sequence will contain pulse-train information paired with their corresponding ground-truth SoC value and can be defined by $D = \left\{{(\psi(1),SoC(1)^*),(\psi(2),SoC(2)^*),...,(\psi(\tau),SoC(\tau)^*)}\right\}$, where $SoC(p)^*$ and $\psi(p)$ are the ideal state-of-charge value and the vector representing the pulse-train input.

FNNs can be summarized by a sequence of matrix multiplication and can be represented by the below composite function.  Let $w_{j,k}^{l}$ denote the weight connection between neuron $j$ in layer $l-1$ and neuron $k$ in layer $l$.  Let $b_{k}^{l}$ and $h_{k}^{l}$ be the bias and the activation function, respectively, of neuron $k$ in layer $l$ .  The hidden layer activations can be computed as follows;
\begin{equation} \label{eq:ForwardPass}
\begin{aligned}
h_{k}^{l}(p) &= \eta\left(\sum_{k}(w_{j,k}^{l}h_{k}^{l-1}(p) + b_{k}^{l})\right)\\
\end{aligned}
\end{equation} 
where,
\begin{equation} \label{eq:ForwardPass}
\begin{aligned}
h_{k}^{l}(p) &= SoC(p) & \text{for $l = L$}\\
\end{aligned}
\end{equation} 
$SoC(p)$ is the estimated state-of-charge for pulse-train $p$.  The nonlinearity used in these networks is called Rectified Linear Units (ReLU) due to its simplicity during the feedforward and backpropagation steps.  The latter is given by;
\begin{equation} \label{eq:nonlinearity}
\begin{aligned}
\eta &= max(0,h)\\
\end{aligned}
\end{equation}
The error signal measuring similarity of the estimated SoC value to the gound-truth value is given by;
\begin{equation} \label{eq:loss}
\begin{aligned}
e(p) &= SoC(p) - SoC^*(p)\\
\end{aligned}
\end{equation}
The a mean absolute error summarizes the performance of the FNN over the entire dataset and is defined by;
\begin{equation} \label{eq:loss}
\begin{aligned}
\mathcal{L} &= \frac{1}{\tau}\sum_{t=0}^\tau\left(e(p)^2\right)
\end{aligned}
\end{equation}
where $\tau$ is the length of the pulse-train.  A forward pass begins when the pulses are fed into the network and is complete when the FNN provides an estimate of the SoC and the over loss is computed.  A full training epoch, $\epsilon$, includes one forward pass and one backward pass; describing the process of tuning the network weights and biases based on the loss function.  This is defined by the following composite function;
\begin{equation} \label{eq:BackwardPass}
\begin{aligned}
m_\epsilon &= \beta_1 m_{\epsilon-1}\nabla \mathcal{L}(w_{\epsilon-1})\\
r_\epsilon &= \beta_2 r_{\epsilon-1}\nabla \mathcal{L}(w_{\epsilon-1})^2\\
\widetilde{m}_\epsilon &= m_\epsilon/(1-{\beta_1}^\epsilon)\\
\widetilde{r}_\epsilon &= r_\epsilon/(1-{\beta_2}^\epsilon)\\
w_\epsilon &= w_{\epsilon-1}-\alpha \frac{\widetilde{m}_\epsilon}{\widetilde{r}_\epsilon - \kappa}, 
\end{aligned}
\end{equation} 
where $\beta_1$ and $\beta_2$ are decay rates set to 0.9 and 0.999, respectively, $\alpha$ is the learning rate and $\kappa$ is a constant term set to $10^{-8}$.

Training of the FNN is done offline and only when  network converges to a lower loss threshold can the networks be applied online.  During online operation, only a forward pass is required in order to estimate SoC.  Backward passes are no longer required once the model is appropriately trained.  FNNs offer an advantage of faster computing time, once trained, since a forward pass is comprised mainly of a sequence of matrix multiplications.

In this paper, TensorFlow \cite{tf2015}, a machine learning framework, is used with a TITAN Xp NVIDIA Graphical Processing Unit (GPU).  The TensorFlow and Keras frameworks provide the ability to prototype neural networks quickly and iterate on various architectures and loss functions. These frameworks also offer automatic gradient computation thereby allowing for a seamless backward computation without any manual intervention.

\begin{figure}
    \centering
    \subfigure[]{
        \includegraphics[height=0.25\textwidth]{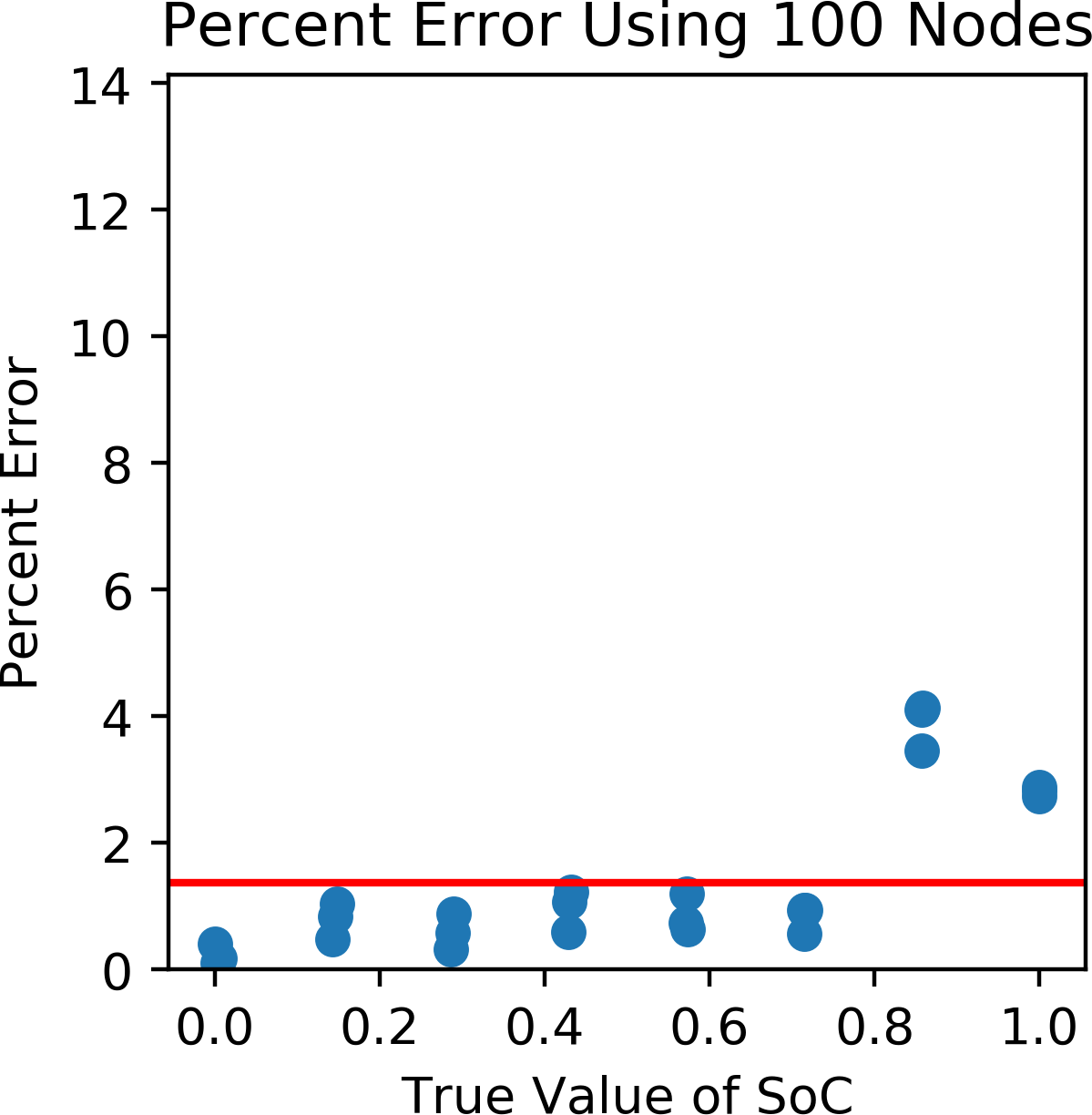}
        \label{fig:I_100_capcheck}
    }
    
    \subfigure[]{
        \includegraphics[height=0.25\textwidth]{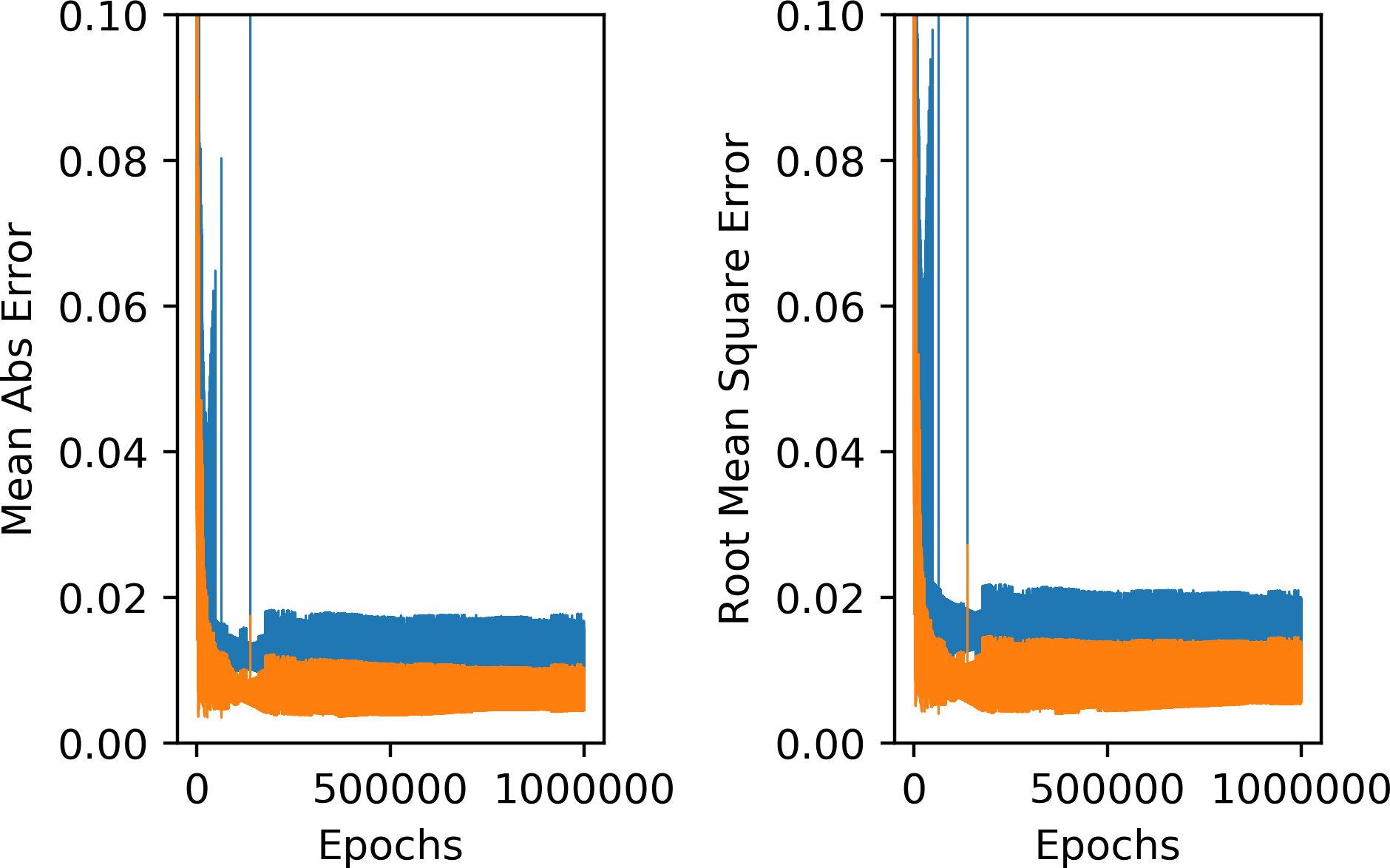}
        \label{fig:V_100_capcheck}
    }
    \caption{Estimation accuracy of a feed-forward neural network reconstructing SoC from a finite-time test sequence with $N$=100 hidden nodes. a) SOC percent error recorded as a function of ground-truth SOC values.  b) Example of training process for FNNs.  MAE and RMSE over training and validation datasets are recorded as a function of training epochs.}
    \label{fig:FNN_100}
\end{figure}

\begin{figure}
    \centering
    \subfigure[]{
        \includegraphics[height=0.25\textwidth]{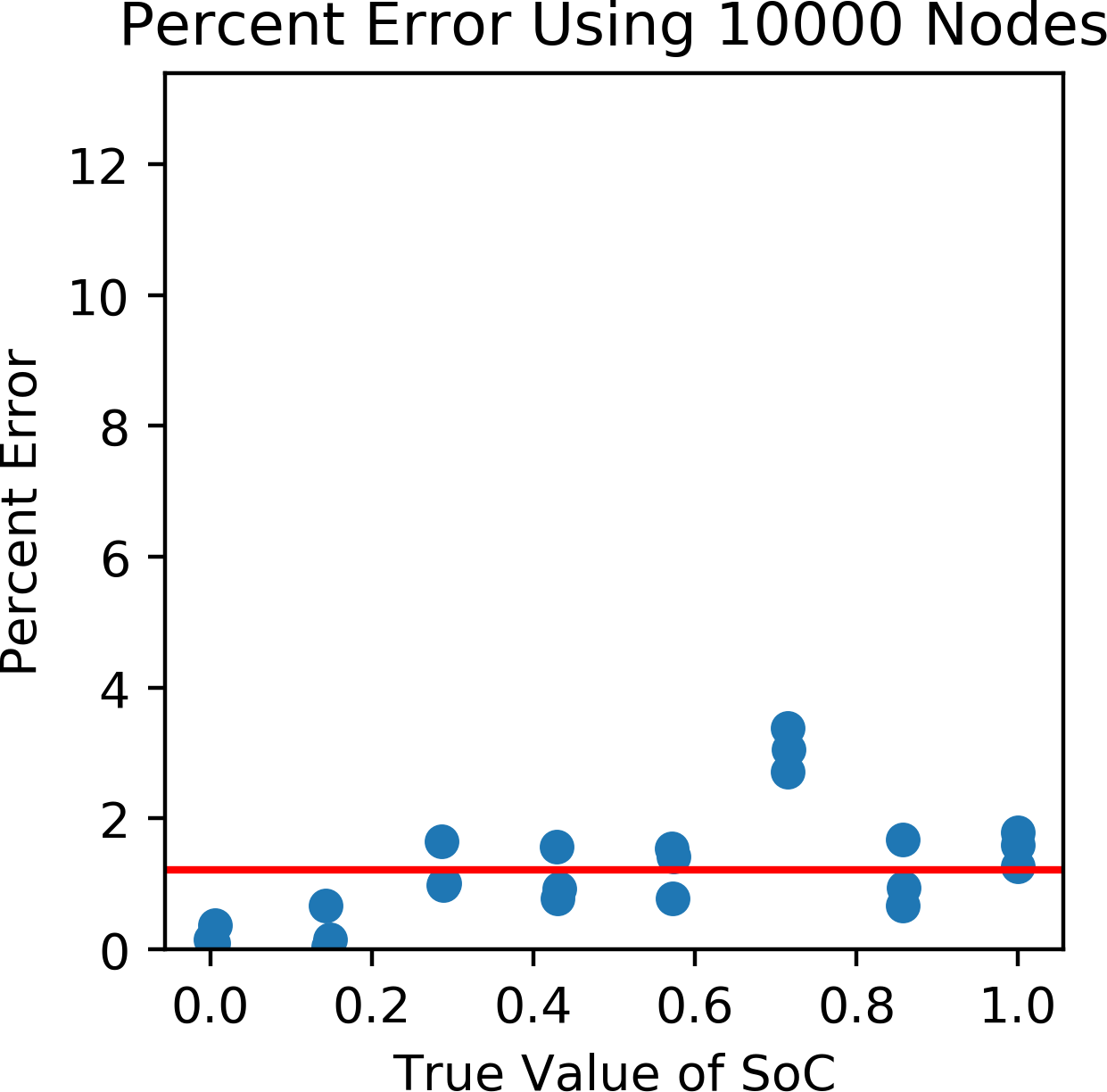}
        \label{fig:I_10000_capcheck}
    }
    
    \subfigure[]{
        \includegraphics[height=0.25\textwidth]{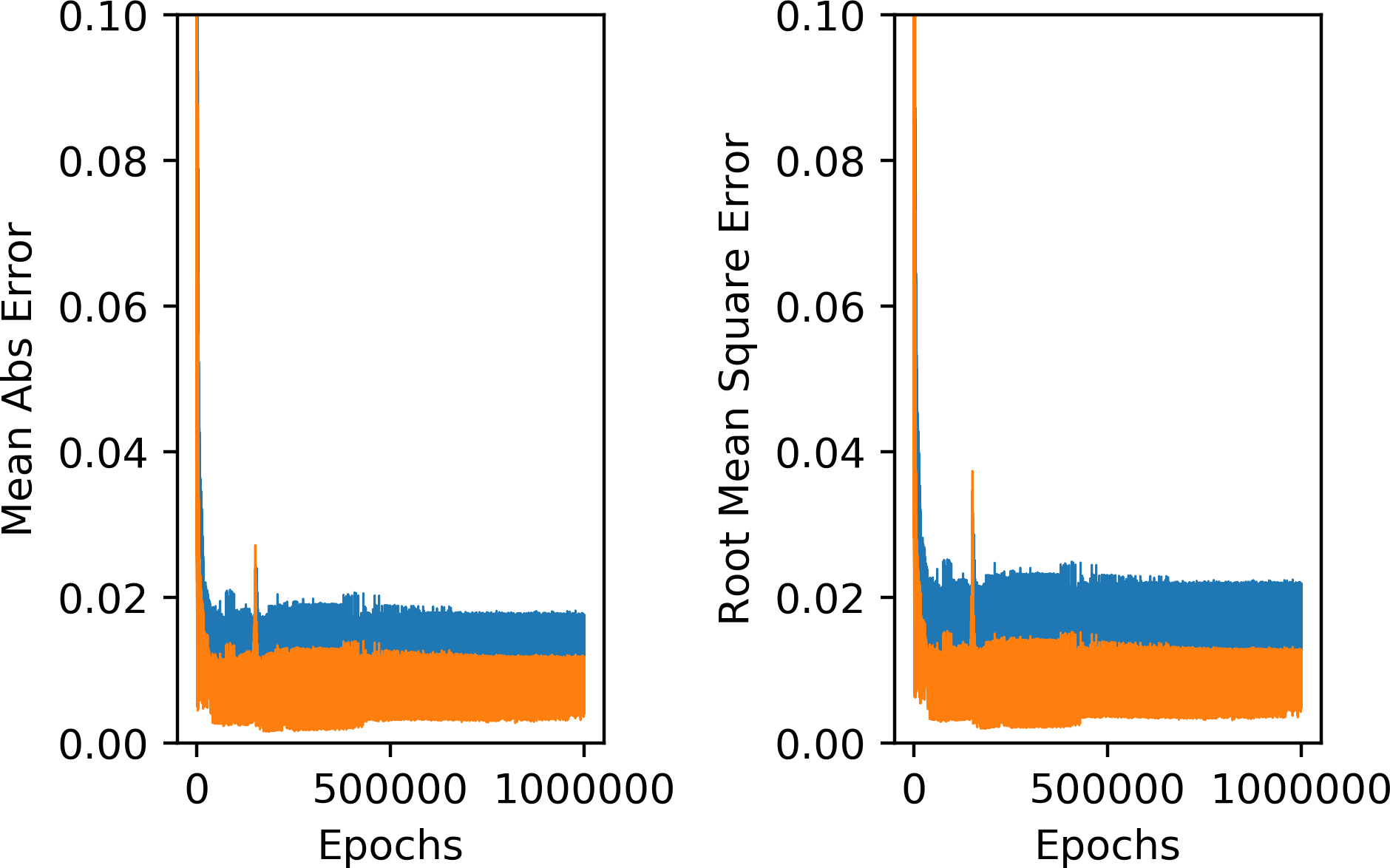}
        \label{fig:V_10000_capcheck}
    }
    \caption{Estimation accuracy of a feed-forward neural network reconstructing SoC from a finite-time test sequence with $N$=10,000 hidden nodes. a) SOC percent error recorded as a function of ground-truth SOC values. b) Example of training process for FNNs.  MAE and RMSE over training and validation datasets are recorded as a function of training epochs.}
    \label{fig:FNN_10000}
\end{figure}

An example of the training and validation process is shown in Fig.~\ref{fig:FNN_100} and Fig.~\ref{fig:FNN_10000}, where the estimation error is shown as a function of the true SOC values or ground-truth values in Fig.~\ref{fig:I_100_capcheck} and Fig.~\ref{fig:I_10000_capcheck}.  The mean error across the entire SOC range is well below 2\% which is quite competitive compared with aforementioned model-based observers.  In Fig.~\ref{fig:V_100_capcheck} and Fig.~\ref{fig:V_10000_capcheck}, the model is trained for 10,000 epochs; training and validation MAE are shown. In this work, training typically spanned 1 to 10 hours depending on the number of epochs chosen.


\section{Application in Electric Vehicles}
\label{sec:pack}
Although training is done on a GPU to capitalize on their parallel computing capability, when applying the FNNs in real world situations, a standard microprocessor can be used since, as mentioned above, the feedforward step comprises of a series of matrix multiplications. 

\begin{figure}[]
	\centering
	\includegraphics[width=0.45\textwidth]{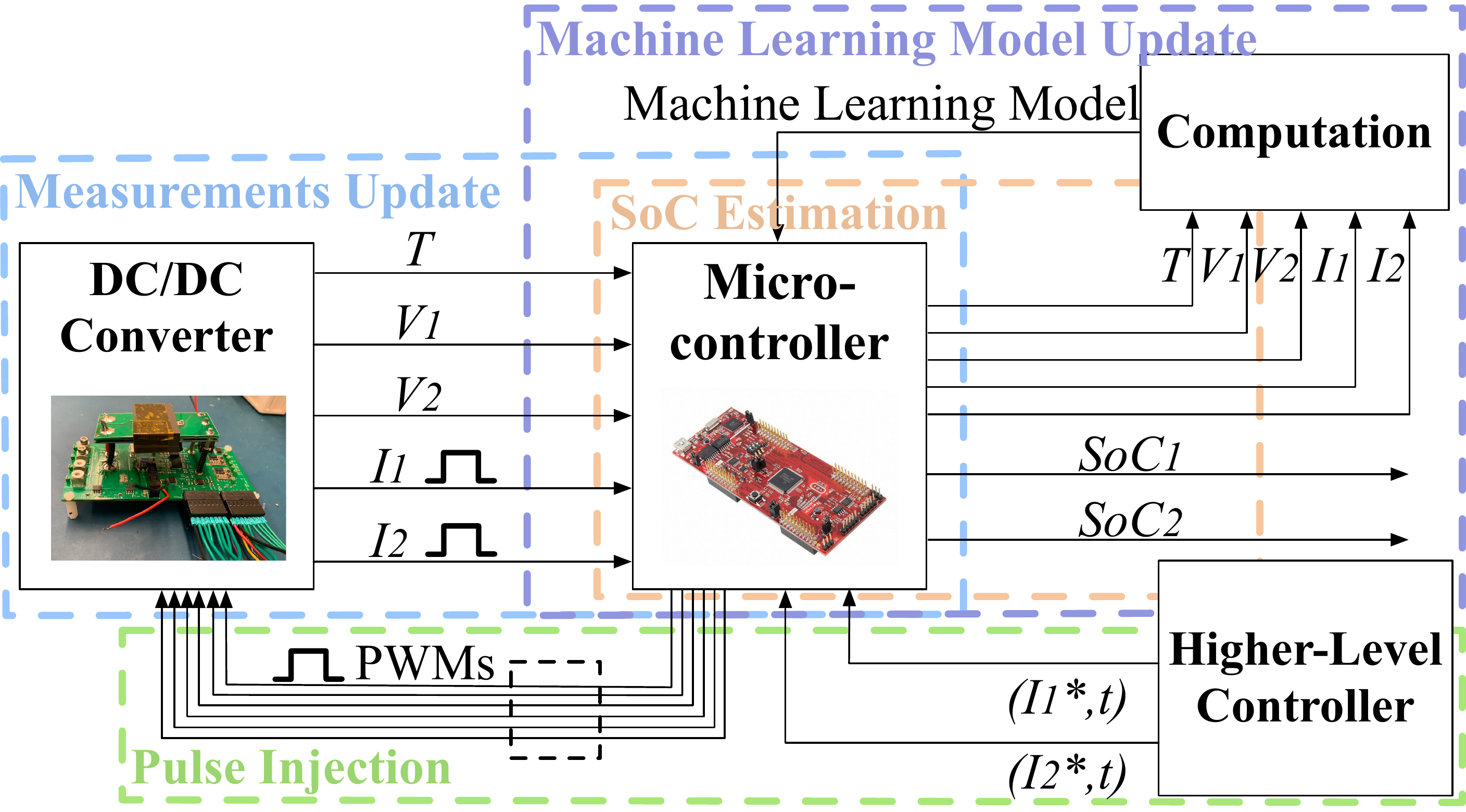}
	\caption{
	Machine learning algorithm implemented in real battery system
	}
	\label{fig:system_diagram}
\end{figure}

A summary of experiment testbench to implement the proposed machine learning algorithm in real battery systems is presented. The diagram of the testbench is shown in Fig. \ref{fig:system_diagram}. Note that the DC/DC converter, illustrated in Fig. \ref{fig:system_diagram}, has already been built. The DC/DC converter is a battery balancing circuit equipped with sensors and necessary computing unit to perform basic BMS functions \cite{Wang2018b,Wang2018c}. 
In addition, the microcontroller has been integrated with the converter to perform reasonable computing work. The entire system consists of (i) the peripheral hardware (the DC/DC converter); (ii) center and localized controllers that actuates the pulse injection, necessary BMS functions and circuit operation; and (iii) powerful computing units that analyze the uploaded data and generate the machine learning model for the use of the microcontroller.


\subsection{Pulse Injection Module}
As the key novelty of the proposed concept, the pulses should be injected to the cells at the right timing with proper amplitude and duration as accurate as in laboratory environment. The higher-level controller which acts like a ‘brain’ of the BMS initially sends commands to the local microcontroller. This command describes the reference currents of the battery cells, specifying amplitude and duration of expected pulses. Once the command is received by the microcontroller, it will generate corresponding PWM pulses and pass them to DC/DC converter to actuate the pulses into the battery cells. Additionally, the property (amplitude, period, etc.) of pulses could be arbitrarily adjusted by properly controlling the converter’s behavior.  

\subsection{Measurements Update Module }
The microcontroller equips high-resolution analog-to-digital converters that translates the analog signals (such as voltage/current measurements) to digital values, such that the computing unit can process them. The essential measurements (cell voltages, currents and temperatures) that are necessary inputs for the machine learning algorithm are captured and updated at the pre-defined sampling rate. Based on the sampling rate, the measurements will be continuously uploaded to the computing unit via communication protocol for further calculations of the machine learning model.

\subsection{Machine Learning Model Update}
During machine learning model update, aforementioned measurements captured by the microcontroller have been transmitted to computation center either through wire or wireless. The machine learning model is iteratively trained and updated using accumulated cell information.
At every pre-defined rate, the updated machine learning model is sent back to the local controllers for corrections of aging side-effects and temperature changes. 

\subsection{SoC Estimation}
As explained previously, the current pulses are injected to the cells through pulse injection module. The corresponding responses from battery cells are recorded by measurement update module to be used as inputs for the SoC estimation using machine learning model. The SoC estimation can be performed based on the measurement with the machine learning model, taking advantage of the simplified matrix multiplications.

Two approaches using pulse injection to augment SoC estimation have been considered as candidates to update SoC estimation in real-time. 1) The machine learning model is continuously operating to obtain the SoC values in real-time. 2) machine learning model only operates at certain moments, for example when the vehicle stops at red light. Between the times when SOC estimation is updated by machine learning, other SoC estimation techniques (such as coulomb counting and EKF) can be applied to estimate SoC for those cost- and computation-constrained applications. 

\begin{figure}[]
	\centering
	\includegraphics[width=0.45\textwidth]{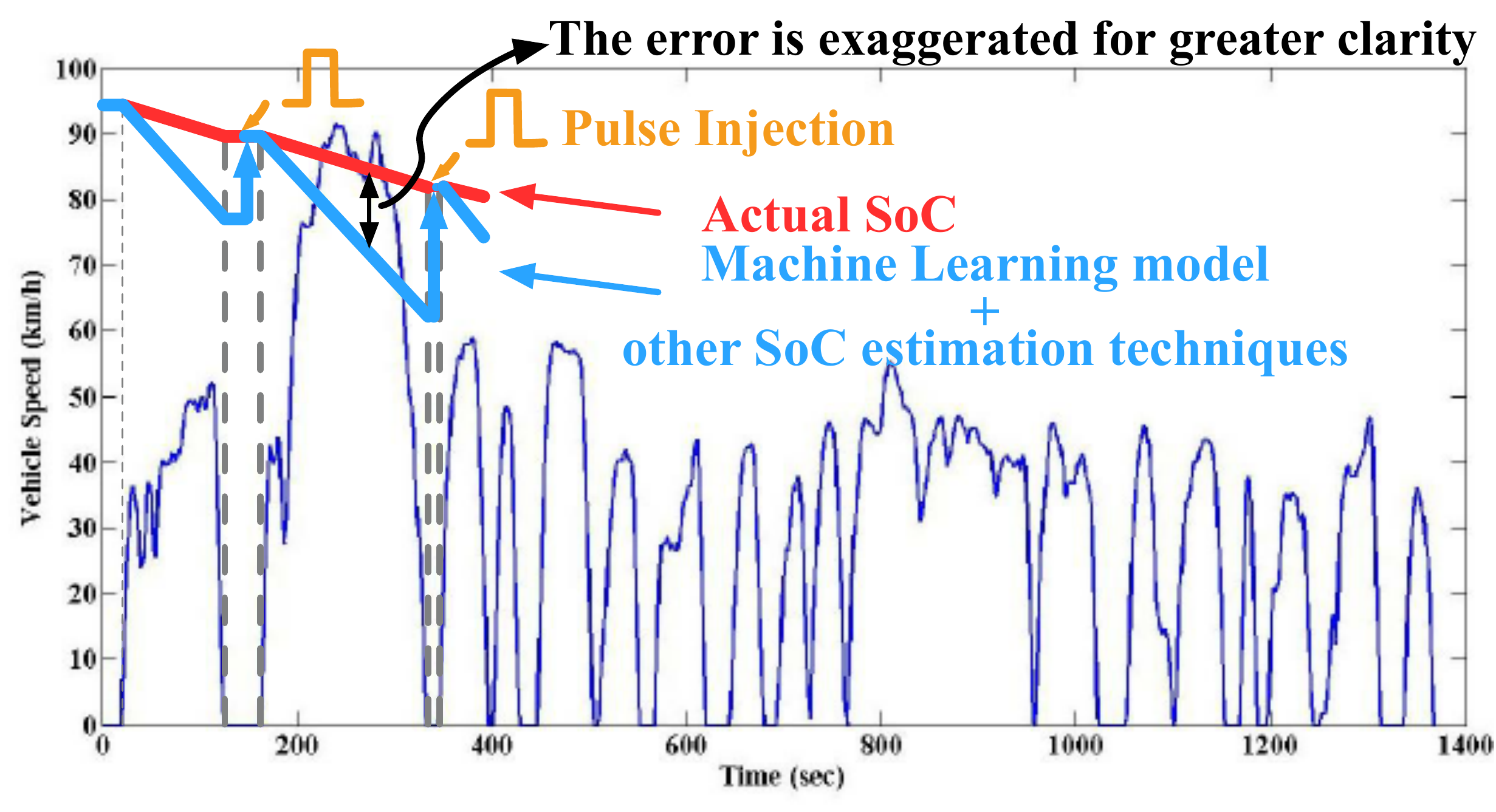}
	\caption{
	SoC estimation in real battery system with UDDS driving cycle \cite{Barlow2009}
	}
	\label{fig:udds}
\end{figure}

Fig. \ref{fig:udds} showcases how the latter method is employed in the actual battery system with UDDS driving cycle. The red line represents the actual SoC. The blue one illustrates the SoC estimation algorithm of the latter approach, where SoC resets at every point when the electric vehicle stops. After the vehicle restarts, other SoC estimation technique (such as coulomb counting or EKF) resumes. As a result, accumulated error in the previous driving period will be eliminated. 

Please note that the error presented between the two SoC curves, in Fig. \ref{fig:udds}, is exaggerated for greater clarity. In practice, the difference between them will be heavily dependent on the SoC estimation strategy adopted in the system and can be reduced significantly by correcting the estimated SoC more regularly. 

\section{Conclusions and Future Work}
\label{sec:Conclusion}
This paper introduces a new strategy to augment the performance of SoC estimation powered by neural networks. A high-fidelity electrochemical battery model is used to validate the concept of the pulse injection and demonstrate that higher current amplitudes contribute to more accurate SoC estimation. As the first batch of cell data is usually acquired from laboratory environment, the testing procedure tailored for the pulse-injection augmentation is discussed and detailed steps are given. The method to construct FNN for mapping the pulse measurements to a ground-truth is provided. By applying FNN to the data, the SoC can be reconstructed within a error boundary of $\pm$2\%. Thanks to the advantage of FNN, a standard microprocessor is capable of running FNNs with just simplified matrix multiplications after the model is trained, which makes real-time computing feasible. Lastly, the experimental validation platform has been demonstrated and explained. By using a BMS-ready balancing circuit that is previously developed, the pulse injection can be integrated into balancing current demands without interfering with driving behaviors.

At the time of writing this paper, the experimental setup has been completed and the experimental tests are undergoing on a scaled battery-pack. The test-bench that is representative of real-world conditions in transportation electrification and the SoC estimation results will be published.
Furthermore, the authors intend to apply the proposed technique to monitor battery aging. In fact, the battery response to a time series is expected to encode a range of information on the internal behavior of the battery. Future research will study finite-time sequences that can reveal general aging information, i.e. the State of Health (SoH), as well as specific aging effects such as active material dissolution, surface layer formation, and atomic structure rearrangement.

\section{Acknowledgement}
This research was undertaken, in part, through funding from the Columbia University Data Science Institute (DSI) Seed Fund Program. It was facilitated by NVIDIA Corporation with the donation of a Titan Xp GPU. We acknowledge computing resources from Columbia University’s Shared Research Computing Facility project, which is supported by NIH Research Facility Improvement grant 1G20RR030893-01, and associated funds from the New York State Empire State Development, Division of Science Technology and Innovation (NYSTAR) Contract C090171, both awarded April 15, 2010. We would also like to thank Robert C. Mohr for his contributions in the experimental setup.


\section*{References}
\printbibliography[heading=none]

@article{Dai2012,
author = {Dai, Yiling and Cai, Long and White, Ralph E},
doi = {10.1149/2.026302jes},
isbn = {0013-4651},
issn = {0013-4651},
journal = {Journal of the Electrochemical Society},
number = {1},
pages = {A182--A190},
title = {{Capacity Fade Model for Spinel LiMn2O4 Electrode}},
url = {http://jes.ecsdl.org/content/160/1/A182.abstract{\%}5Cnhttp://jes.ecsdl.org/cgi/doi/10.1149/2.026302jes},
volume = {160},
year = {2012}
}

@article{Wang2016,
author = {Wang, Zong-Kai and Shu, Jie and Zhu, Qian-Cheng and Cao, Bo-Yu and Chen, Hui and Wu, Xue-Yan and Bartlett, Bart M. and Wang, Kai-Xue and Chen, Jie-Sheng},
doi = {10.1016/j.jpowsour.2016.01.005},
issn = {03787753},
journal = {Journal of Power Sources},
month = {mar},
pages = {426--434},
title = {{Graphene-nanosheet-wrapped LiV 3 O 8 nanocomposites as high performance cathode materials for rechargeable lithium-ion batteries}},
url = {http://linkinghub.elsevier.com/retrieve/pii/S0378775316300052},
volume = {307},
year = {2016}
}

@article{Xiong2017,
author = {Xiong, Rui and Cao, Jiayi and Yu, Quanqing and Sun, Fengchun},
doi = {10.1109/ACCESS.2017.2780258},
journal = {IEEE Access},
pages = {1832--1843},
title = {{SPECIAL SECTION ON BATTERY ENERGY STORAGE AND MANAGEMENT SYSTEMS Critical Review on the Battery State of Charge Estimation Methods for Electric Vehicles}},
url = {http://www.ieee.org/publications{\_}standards/publications/rights/index.html},
volume = {6},
year = {2017}
}

@article{Brady2016,
author = {Brady, Nicholas W and Zhang, Qing and Knehr, K W and Liu, Ping and Marschilok, Amy C and Takeuchi, Kenneth J and Takeuchi, Esther S and West, Alan C},
doi = {10.1149/2.0341614jes},
journal = {Journal of the Electrochemical Society},
number = {14},
pages = {2890--2898},
title = {{Discharge, Relaxation, and Charge Model for the Lithium Trivanadate Electrode: Reactions, Phase Change, and Transport}},
volume = {163},
year = {2016}
}

@article{Brady2018,
author = {Brady, Nicholas W. and Zhang, Qing and Bruck, Andrea and Bock, David C. and Gould, Christian Alexander and Marschilok, Amy C. and Takeuchi, Kenneth and Takeuchi, Esther and West, Alan C.},
doi = {10.1149/2.1291802jes},
issn = {0013-4651},
journal = {Journal of The Electrochemical Society},
month = {feb},
number = {2},
pages = {A371--A379},
publisher = {The Electrochemical Society},
title = {{Operando Study of LiV3O8 Cathode: Coupling EDXRD Measurements to Simulations}},
url = {http://jes.ecsdl.org/lookup/doi/10.1149/2.1291802jes},
volume = {165},
year = {2018}
}

@techreport{Barlow2009,
author = {Barlow, Tim J and Latham, S and McCrae, Ian S and Boulter, Paul G},
institution = {The Future of Transport},
pages = {284},
title = {{A reference book of driving cycles for use in the measurement of road vehicle emissions}},
year = {2009}
}

@article{Charkhgard2010,
author = {Charkhgard, Mohammad and Farrokhi, Mohammad},
doi = {10.1109/TIE.2010.2043035},
issn = {02780046},
journal = {IEEE Transactions on Industrial Electronics},
number = {12},
pages = {4178--4187},
publisher = {IEEE},
title = {{State-of-charge estimation for lithium-ion batteries using neural networks and EKF}},
volume = {57},
year = {2010}
}

@article{Wang2015,
author = {Wang, Weizhong and Ye, Jin and Malysz, Pawel and Yang, Hong and Emadi, Ali},
isbn = {9781467367417},
journal = {IEEE Transportation Electrification Conference},
pages = {1--7},
title = {{Sensitivity Analysis of Kalman filter Based Capacity Estimation for Electric Vehicles}},
year = {2015}
}

@inproceedings{Wang2016c,
author = {Wang, Weizhong and Wang, Deqiang and Wang, Xiao and Li, Tongrui and Ahmed, Ryan and Habibi, Saeid and Emadi, Ali},
booktitle = {IEEE Transportation Electrification Conference and Expo (ITEC), Dearborn, MI},
pages = {1--6},
title = {{Comparison of Kalman Filter-based State of Charge Estimation Strategies for Li-Ion Batteries}},
year = {2016}
}

@article{Waag2013c,
author = {Waag, Wladislaw and Fleischer, Christian and Sauer, Dirk Uwe},
doi = {10.1016/j.jpowsour.2013.05.111},
isbn = {0378-7753},
issn = {03787753},
journal = {Journal of Power Sources},
pages = {548--559},
publisher = {Elsevier B.V},
title = {{Adaptive on-line prediction of the available power of lithium-ion batteries}},
url = {http://dx.doi.org/10.1016/j.jpowsour.2013.05.111},
volume = {242},
year = {2013}
}

@article{Ng2009,
author = {Ng, Kong Soon and Moo, Chin Sien and Chen, Yi Ping and Hsieh, Yao Ching},
doi = {10.1016/j.apenergy.2008.11.021},
isbn = {9781424424900},
issn = {03062619},
journal = {Applied Energy},
number = {9},
pages = {1506--1511},
publisher = {Elsevier Ltd},
title = {{Enhanced coulomb counting method for estimating state-of-charge and state-of-health of lithium-ion batteries}},
url = {http://dx.doi.org/10.1016/j.apenergy.2008.11.021},
volume = {86},
year = {2009}
}

@article{Baronti2011,
author = {Baronti, F. and Fantechi, G. and Fanucci, L. and Leonardi, E. and Roncella, R. and Saletti, R. and Saponara, S.},
isbn = {9788070439876},
issn = {18037232},
journal = {International Conference on Applied Electronics},
number = {1},
pages = {29--34},
publisher = {IEEE},
title = {{State-of-charge estimation enhancing of lithium batteries through a temperature-dependent cell model}},
volume = {0},
year = {2011}
}

@phdthesis{Wang2016b,
author = {Wang, Weizhong},
pages = {xx, 149},
school = {Master's Thesis, Department of Electrical and Computer Engineering, McMaster University, Hamilton, ON, Canada},
title = {{Modeling , Estimation and Benchmarking of Lithium Ion Electric Bicycle Battery}},
url = {http://hdl.handle.net/11375/20293},
year = {2016}
}

@article{Current,
author = {Current, Saturation},
pages = {1--2},
title = {{Using Standard Transformers in Multiple Applications}}
}

@article{Xia2018,
author = {Xia, Bizhong and Lao, Zizhou and Zhang, Ruifeng and Tian, Yong and Chen, Guanghao and Sun, Zhen and Wang, Wei and Sun, Wei and Lai, Yongzhi and Wang, Mingwang and Wang, Huawen},
doi = {10.3390/en11010003},
issn = {19961073},
journal = {Energies},
number = {1},
title = {{Online parameter identification and state of charge estimation of lithium-ion batteries based on forgetting factor recursive least squares and nonlinear Kalman filter}},
volume = {11},
year = {2018}
}

@article{Roscher2011,
archivePrefix = {arXiv},
arxivId = {ID 984320},
author = {Roscher, Michael a. and Bohlen, Oliver and Vetter, Jens},
doi = {10.4061/2011/984320},
eprint = {ID 984320},
issn = {2090-3537},
journal = {International Journal of Electrochemistry},
pages = {1--6},
title = {{OCV Hysteresis in Li-Ion Batteries including Two-Phase Transition Materials}},
url = {http://www.hindawi.com/journals/ijelc/2011/984320/},
volume = {2011},
year = {2011}
}

@article{Chemali2018,
author = {Chemali, Ephrem and Kollmeyer, Phillip J. and Preindl, Matthias and Emadi, Ali},
doi = {10.1016/j.jpowsour.2018.06.104},
issn = {03787753},
journal = {Journal of Power Sources},
number = {August},
pages = {242--255},
publisher = {Elsevier},
title = {{State-of-charge estimation of Li-ion batteries using deep neural networks: A machine learning approach}},
url = {https://doi.org/10.1016/j.jpowsour.2018.06.104},
volume = {400},
year = {2018}
}

@article{Du2014,
author = {Du, Jiani and Liu, Zhitao and Wang, Youyi},
doi = {10.1016/j.conengprac.2013.12.014},
issn = {09670661},
journal = {Control Engineering Practice},
number = {1},
pages = {11--19},
publisher = {Elsevier},
title = {{State of charge estimation for Li-ion battery based on model from extreme learning machine}},
url = {http://dx.doi.org/10.1016/j.conengprac.2013.12.014},
volume = {26},
year = {2014}
}

@article{Waag2014,
author = {Waag, Wladislaw and Fleischer, Christian and Sauer, Dirk Uwe},
doi = {10.1016/j.jpowsour.2014.02.064},
isbn = {0378-7753},
issn = {03787753},
journal = {Journal of Power Sources},
pages = {321--339},
publisher = {Elsevier B.V},
title = {{Critical review of the methods for monitoring of lithium-ion batteries in electric and hybrid vehicles}},
url = {http://dx.doi.org/10.1016/j.jpowsour.2014.02.064},
volume = {258},
year = {2014}
}

@article{Idaho2010,
author = {Idaho, The and National, Energy},
number = {December},
title = {{U . S . Department of Energy Vehicle Technologies Program Battery Test Manual For Plug-In Hybrid Electric Vehicles}},
year = {2010}
}

@inproceedings{Wang2018b,
author = {Wang, Weizhong and Preindl, Matthias},
booktitle = {IEEE Applied Power Electronics Conference and Exposition (APEC)},
doi = {10.1109/APEC.2018.8341582},
isbn = {978-1-5386-1180-7},
pages = {3340--3345},
title = {{Modeling and control of a dual cell link for battery-balancing auxiliary power modules}},
year = {2018}
}

@article{Belhani2013,
author = {Belhani, Ahmed and M'Sirdi, Nacer K. and Naamane, Aziz},
doi = {10.1016/j.egypro.2013.11.038},
issn = {18766102},
journal = {Energy Procedia},
pages = {377--386},
title = {{Adaptive sliding mode observer for estimation of state of charge}},
volume = {42},
year = {2013}
}

@article{Plett2005,
author = {Plett, G},
isbn = {9781632668394},
journal = {In Proceedings of the 21st Electric Vehicle Symposium (EVS21), Monaco},
pages = {1--12},
title = {{Dual and Joint EKF for Simultaneous SOC and SOH Estimation}},
url = {http://mocha-java.uccs.edu/dossier/RESEARCH/2005evs21b-.pdf},
year = {2005}
}

@inproceedings{Wang2018c,
author = {Wang, Weizhong and Preindl, Matthias},
booktitle = {IEEE Transportation Electrification Conference and Expo (ITEC)},
doi = {10.1109/ITEC.2018.8450218},
isbn = {978-1-5386-3048-8},
issn = {1048-2334},
pages = {898--903},
title = {{Design and Implementation of a Dual Cell Link for Battery-Balancing Auxiliary Power Modules}},
year = {2018}
}

@article{Ning2016,
author = {Ning, Bo and Xu, Jun and Cao, Binggang and Wang, Bin and Xu, Guangcan},
doi = {10.1016/j.egypro.2016.06.088},
issn = {18766102},
journal = {Energy Procedia},
pages = {619--626},
publisher = {Elsevier B.V.},
title = {{A sliding mode observer SOC estimation method based on parameter adaptive battery model}},
url = {http://dx.doi.org/10.1016/j.egypro.2016.06.088},
volume = {88},
year = {2016}
}

@article{Plett2004c,
author = {Plett, Gregory L.},
doi = {10.1016/j.jpowsour.2004.02.033},
isbn = {0378-7753},
issn = {03787753},
journal = {Journal of Power Sources},
pages = {277--292},
title = {{Extended Kalman filtering for battery management systems of LiPB-based HEV battery packs - Part 3. State and parameter estimation}},
volume = {134},
year = {2004}
}

@article{Liu2016,
author = {Liu, Congzhi and Liu, Weiqun and Wang, Lingyan and Hu, Guangdi and Ma, Luping and Ren, Bingyu},
doi = {10.1016/j.jpowsour.2016.03.112},
issn = {03787753},
journal = {Journal of Power Sources},
pages = {1--12},
publisher = {Elsevier B.V},
title = {{A new method of modeling and state of charge estimation of the battery}},
url = {http://dx.doi.org/10.1016/j.jpowsour.2016.03.112},
volume = {320},
year = {2016}
}

@article{Sepasi2014,
author = {Sepasi, Saeed and Ghorbani, Reza and Liaw, Bor Yann},
doi = {10.1016/j.jpowsour.2013.12.093},
issn = {03787753},
journal = {Journal of Power Sources},
pages = {368--376},
publisher = {Elsevier B.V},
title = {{Improved extended Kalman filter for state of charge estimation of battery pack}},
url = {http://dx.doi.org/10.1016/j.jpowsour.2013.12.093},
volume = {255},
year = {2014}
}

@article{He2012,
author = {He, Hongwen and Zhang, Xiaowei and Xiong, Rui and Xu, Yongli and Guo, Hongqiang},
doi = {10.1016/j.energy.2012.01.009},
issn = {03605442},
journal = {Energy},
number = {1},
pages = {310--318},
publisher = {Elsevier Ltd},
title = {{Online model-based estimation of state-of-charge and open-circuit voltage of lithium-ion batteries in electric vehicles}},
url = {http://dx.doi.org/10.1016/j.energy.2012.01.009},
volume = {39},
year = {2012}
}

@article{Chemali2016,
author = {Chemali, Ephrem and Preindl, Matthias and Malysz, Pawel and Emadi, Ali},
doi = {10.1109/JESTPE.2016.2566583},
isbn = {2168-6777 VO - PP},
issn = {2168-6777},
journal = {IEEE Journal of Emerging and Selected Topics in Power Electronics},
number = {3},
pages = {1117--1134},
title = {{Electrochemical and Electrostatic Energy Storage and Management Systems for Electric Drive Vehicles: State-of-the-Art Review and Future Trends}},
url = {http://ieeexplore.ieee.org/lpdocs/epic03/wrapper.htm?arnumber=7467413},
volume = {4},
year = {2016}
}

@Article{Chemali2017,
  author       = {Ephrem Chemali and Phil Kollmeyer and Matthias Preindl and Ryan Ahmed and Ali Emadi},
  title        = {Long Short-Term Memory-Networks for Accurate State of Charge Estimation of Li-ion Batteries},
  journaltitle = {IEEE Transactions on Industrial Electronics},
  date         = {2018},
  volume       = {65},
  pages        = {6730 - 6739},
  doi          = {10.1109/TIE.2017.2787586},
  author+an    = {1=advisee; 3=myself},
}

@online{BatteryU,
author = {{Battery University}},
title = {{Types of Lithium-ion Batteries}},
url = {https://batteryuniversity.com/learn/article/types_of_lithium_ion},
urldate = {2019-08-25},
year = {2019}
}

@article{Zou,
author = {Zou, Yuan and Hu, Xiaosong and Ma, Hongmin and Li, Shengbo Eben},
journal = {Journal of Power Sources},
pages = {793--803},
title = {{Combined State of Charge and State of Health estimation over lithium-ion battery cell cycle lifespan for electric vehicles}},
year = {2015},
volume = {273},
}

@article{Kri2012,
author = {Krizhevsky, Alex and Sutskever, Ilya and Hinton, Geoffrey E},
isbn = {9781627480031},
issn = {10495258},
journal = {Advances In Neural Information Processing Systems},
pages = {1--9},
title = {{ImageNet Classification with Deep Convolutional Neural Networks}},
year = {2012}
}

@ARTICLE{Hinton2012,
author={G. Hinton and L. Deng and D. Yu and G. E. Dahl and A. r. Mohamed and N. Jaitly and A. Senior and V. Vanhoucke and P. Nguyen and T. N. Sainath and B. Kingsbury},
journal={IEEE Signal Processing Magazine},
title={Deep Neural Networks for Acoustic Modeling in Speech Recognition: The Shared Views of Four Research Groups},
year={2012},
volume={29},
number={6},
pages={82-97},
ISSN={1053-5888},
month={Nov},}

@article{tf2015,
	title={TensorFlow: Large-scale machine learning on heterogeneous systems, 2015. URL h ttp},
	author={Abadi, Mart{\i}n and Agarwal, Ashish and Barham, Paul and Brevdo, Eugene and Chen, Zhifeng and Citro, Craig and Corrado, Greg S and Davis, Andy and Dean, Jeffrey and Devin, Matthieu and others},
	journal={Software available from tensorflow.org}
}

@article{ciresan2012,
  title={Multi-column deep neural network for traffic sign classification},
  author={Cire{\c{s}}An, Dan and Meier, Ueli and Masci, Jonathan and Schmidhuber, J{\"u}rgen},
  journal={Neural Networks},
  volume={32},
  pages={333--338},
  year={2012},
  publisher={Elsevier}
}

@article{Ma2015,
author = {Ma, Junshui and Sheridan, Robert P. and Liaw, Andy and Dahl, George E. and Svetnik, Vladimir},
title = {Deep Neural Nets as a Method for Quantitative Structure–Activity Relationships},
journal = {Journal of Chemical Information and Modeling},
volume = {55},
number = {2},
pages = {263-274},
year = {2015}
}




\end{document}